\documentclass[aps, prm,twocolumn,superscriptaddress, nofootinbib, showpacs]{revtex4-1}

\usepackage{graphicx}
\usepackage{dcolumn}
\usepackage{dcolumn}
\usepackage{textgreek}
\usepackage{textcomp}
\usepackage{adjustbox}
\usepackage{multirow}

\def\TSOHO{\textit{TM}$_3$(SeO$_3$)$_3$H$_2$O}
\def\NSOHO{Ni$_3$(SeO$_3$)$_3$H$_2$O}
\def\MSOHO{Mn$_3$(SeO$_3$)$_3$H$_2$O}
\def\CSOHO{Co$_3$(SeO$_3$)$_3$H$_2$O}

\def\Tn{$T_N$}

\def\degrees{$^{\circ}$}

\def\Asq{\text{\AA}\textsuperscript{2}}

\def\Pob{$P\overline{1}$}

\begin{document}

\onecolumngrid
Notice of Copyright This manuscript has been authored by UT-Battelle, LLC under Contract No. DE-AC05-00OR22725 with the U.S. Department of Energy. The United States Government retains and the publisher, by accepting the article for publication, acknowledges that the United States Government retains a non-exclusive, paid-up, irrevocable, world-wide license to publish or reproduce the published form of this manuscript, or allow others to do so, for United States Government purposes. The Department of Energy will provide public access to these results of federally sponsored research in accordance with the DOE Public Access Plan (http://energy.gov/downloads/doe-public-access-plan).
\twocolumngrid

\clearpage

\preprint{APS/123-QED}

\title{Tunable magnetic order in low-symmetry SeO$_3$ ligand linked \textit{TM}$_3$(SeO$_3$)$_3$H$_2$O (\textit{TM} = Mn, Co and Ni) compounds}

\author{K.M. Taddei}
\thanks{These authors contributed equally}
\affiliation{Neutron Scattering Division, Oak Ridge National Laboratory, Oak Ridge, TN 37831}
\author{L.D. Sanjeewa}
\thanks{These authors contributed equally}
\affiliation{Materials Science and Technology Division, Oak Ridge National Laboratory, Oak Ridge, TN 37831}
\author{J. Xing}
\thanks{These authors contributed equally}
\affiliation{Materials Science and Technology Division, Oak Ridge National Laboratory, Oak Ridge, TN 37831}
\author{Q. Zhang}
\affiliation{Neutron Scattering Division, Oak Ridge National Laboratory, Oak Ridge, TN 37831}
\author{D. Parker}
\affiliation{Materials Science and Technology Division, Oak Ridge National Laboratory, Oak Ridge, TN 37831}
\author{A. Podleznyak}
\affiliation{Neutron Scattering Division, Oak Ridge National Laboratory, Oak Ridge, TN 37831}
\author{C. dela Cruz}
\affiliation{Neutron Scattering Division, Oak Ridge National Laboratory, Oak Ridge, TN 37831}
\author{A.S. Sefat}
\affiliation{Materials Science and Technology Division, Oak Ridge National Laboratory, Oak Ridge, TN 37831}

\date{\today}

\begin{abstract}

Recently frustrated magnetic materials have once again captured the condensed matter community's interests due to renewed evidence of being the best route to achieve quantum spin-liquid type physics. Generally, one has two strategies to achieve magnetic frustration: through geometric means or through interactions with different requirements and length scales. As the former leads to much simpler theoretical treatments it is generally favored and so magnetic sublattices with geometric frustration are sought after. One approach to finding such lattices is to design them chemically by using non-magnetic linker ligands. Here we report on the magnetic properties of one such family of materials, the transition metal (\textit{TM}) selenite hydrate compounds with chemical formula \TSOHO . These materials link highly distorted \textit{TM}O$_6$ octahedra via non-magnetic [SeO$_3$]$^{2+}$ linkers. Using \textit{TM} = Mn, Co and Ni we report on the structural effects of changing the \textit{TM} site and how they may influence the magnetic structure. Using magnetic susceptibility and neutron powder diffraction we identify low temperature magnetic transitions for all three compounds characterized by the onset of long-range antiferromagnetic order with moderate frustration indexes. Consideration of the magnetic structures reveal that the magnetic order is sensitive to the \textit{TM} site ion and is tunable as it is changed - especially from Mn to Co - with changes in both the moment direction and the ordering vector. Field dependent measurements of the susceptibility and heat capacity reveal metamagnetic transitions in both \MSOHO\ and \CSOHO\ indicating nearby magnetic ground states accessible under relatively small applied fields. Density functional theory calculations broadly confirm these results, showing both a sensitivity of the magnetic structure to the \textit{TM} and its local environment. Although no spin liquid behavior is achieved, these results suggest the fruitfulness of such synthesis philosophies and encourage future work to engender higher frustration in these materials via doping, field, pressure or larger linker ligands.

\end{abstract}


\maketitle


\section{\label{sec:intro}Introduction}

Recently the search for orders beyond the Landau paradigm has garnered much interest as such orders are predicted to offer both novel physics and a huge potential for quantum device design. \cite{Chiu2016} Materials showing topological order are characterized not by symmetry breaking but by changes in topological invariants and so have resulting orders which are robust to many of the conditions which otherwise disrupt order, excitations or thermal/transport properties. \cite{Wen2017} Furthermore, some such orders host fractionalized quasiparticles which have novel statistics and allow for behaviors outside of the Fermion/Boson particle description paradigm. \cite{Arovas1984}

While several methods have emerged to find such order, one particularly fruitful pathway (at least for some classes of topological materials) has been by realizing frustrated magnetism through competing magnetic interactions. \cite{Balents2010, Castelnovo2008, Gardner2010} Materials can host competing interactions either through specific geometric arrangements of the magnetic sublattice, magnetic Hamiltonians with multiple interaction terms with different preferred couplings or length scales or combinations of both. \cite{Ramirez1994,Gardner2010} For instance, when spin carriers are located on special lattices based on corner or edge shared triangles such as Kagom\'e and triangular lattices, spin frustration will be present and three-dimensional long-range magnetic order tends to be suppressed, especially for a spin-$\frac{1}{2}$ value which can enhance quantum fluctuations. \cite{Derrida1978,Yan2011a}

Many theoretical studies predict materials with these lattices will exhibit exotic ground states due to the competing antiferromagnetic (AFM) interactions between neighboring spins, such as spin ice, spin glass, and spin liquid. \cite{Gardner2010,Mila2000,Han2012} Of special interest is finding spin-$\frac{1}{2}$ systems which crystallize in low dimensional sublattices. Such arrangements combine quantum confinement effects with the potential for strong quantum fluctuations as well as reduce perturbing interactions from extended magnetic lattices which can help stabilize long range order. \cite{Starykh2015}  In experimental work, the key is to obtain a structurally and chemically clean compound for a model study because spin frustration in these systems is often eliminated by lattice imperfections, next-nearest-neighbor super-exchange and anisotropy, resulting in long-range magnetic order of the traditional type. \cite{Willans2010,Henley1989}

One controllable route to generate geometric magnetic frustration, is through using non-magnetic bridging ligands such as [VO$_4$]$^{3-}$, [PO$_4$]$^{3-}$, [GeO$_4$]$^{4-}$ and [SeO$_3$]$^{2-}$ which can link magnetic lattices and give rise to a variety of structures \cite{Sanjeewa2016, Garlea2019,Zhang2019a,Achary2017,Ren2008a,Harrison1999,Mcmanus1991,Wildner1991,Larranaga2002}. The hope of using such a route to generate potentially frustrated magnetic materials is to gain more control and design magnetic sublattices from the outset rather than search for frustrated physics in existing compounds. In this study, we have attempted to use [SeO$_3$]$^{2-}$ as our non-magnetic linker and have successfully synthesized a novel series of transition metal (\textit{TM}) selenite hydrate compounds stoichiometry \TSOHO\ and oxidation state $TM^{2+}$. These compounds form low symmetry structures with \textit{TM} layers linked into three dimensional (3D) magnetic lattice and have a useful receptivity to different \textit{TM} site ions. Furthermore, the use of linker ligands creates a complex network of \textit{TM}\textemdash O\textemdash \textit{TM} bonds with numerous distinct \textit{TM} sites which may allow multiple competing pathways for both direct and superexchange interactions.

In this paper we present the results of comprehensive magnetic, transport and diffraction measurements on \MSOHO , \CSOHO\ and \NSOHO\ to search for signs of frustrated magnetism. Furthermore, we study how the variability of the \textit{TM} site may allow for tuning of the magnetic properties finding different magnetic structures for each compound. In the $TM^{2+}$ valence, Co is of special interest as in the low spin configuration it realizes a spin-$\frac{1}{2}$ state and so has promise for achieving the sought after quantum phases. Overall we find evidence of moderate frustration in these three materials with all of them having a long range ordered magnetic ground state. Characterization of this order reveals a general sensitivity of the stabilized magnetic structure to the \textit{TM} site. Additionally, field dependent work suggests low field metamagnetic transitions in both \MSOHO\ and \CSOHO , encouraging future work. While our results show long range order in all compounds, several behaviors of \CSOHO\ suggest it has competing magnetic ground states and may be easily tuned between them.  


\begin{figure}
	\includegraphics[width=\columnwidth]{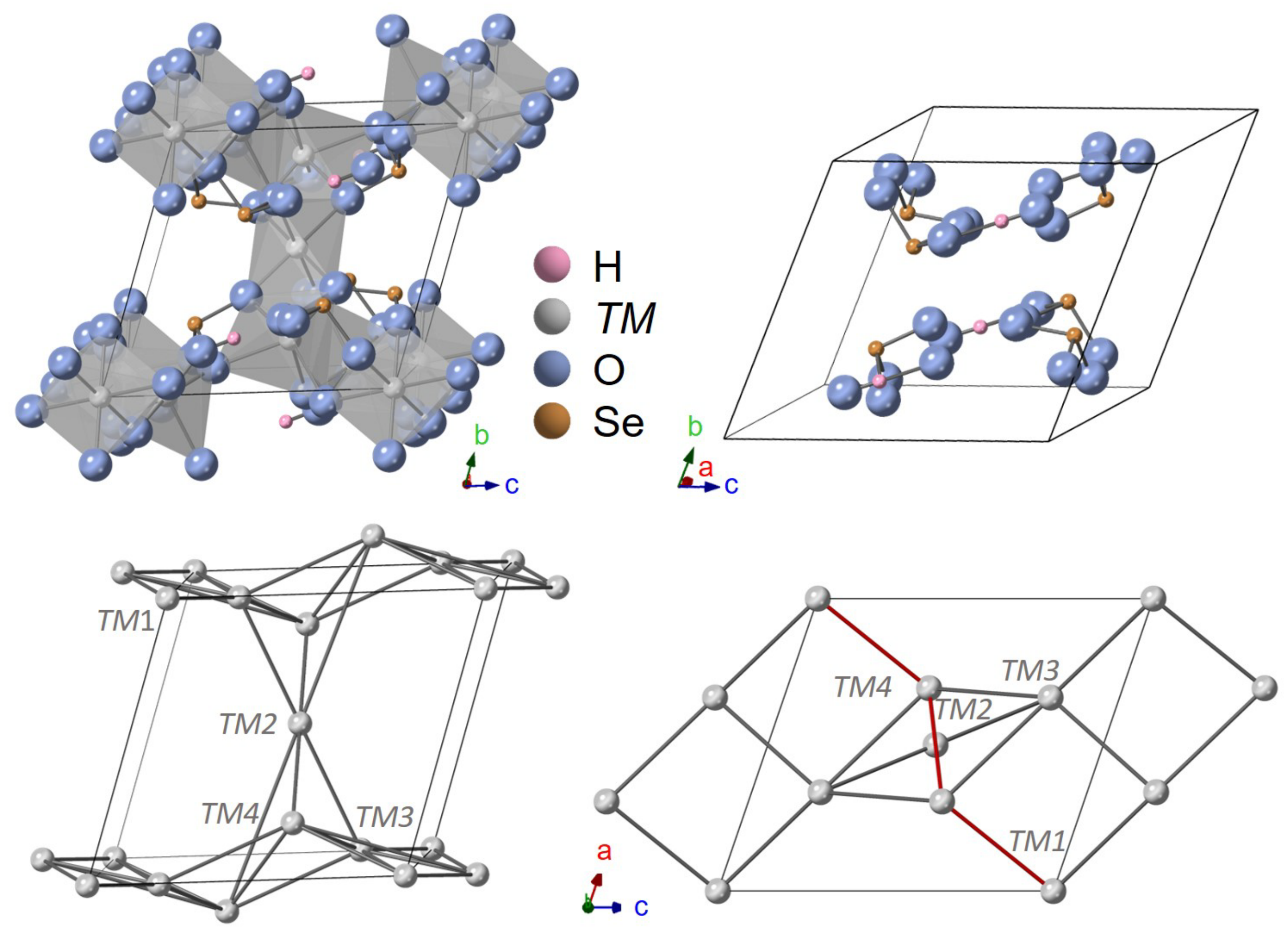}
	\caption{\label{fig:one} Crystal structure of the \TSOHO\ compounds viewed (a) along the \textit{a} lattice direction. (b) Unit cell with \textit{TM} site removed showing the SeO$_3$ linkers and H$_2$O positions. View of \textit{TM} sublattice (c) along the \textit{a} and (d) \textit{b} lattice directions. In panel (d) the red highlighted bonds indicate the \textit{TM}1-\textit{TM}4 quasi-one-dimensional sublattice. Images were generated with CrystalMaker.}	
\end{figure}

\section{\label{sec:methods} Experimental Methods}

\subsection{\label{subsec:synthesis} Synthesis}

A low-temperature hydrothermal technique was employed to synthesize the single crystals of \TSOHO\ (\textit{TM} = Mn, Co, Ni). All the reactions were performed in 23 mL poly(tetrafluroethylene)-lined pressure vessels (Parr instruments) at 220 \degrees C for 2 weeks. To synthesize single crystals of each phase, a mixture of K$_2$CO$_3$ (Aldrich, 99\%), \textit{TM}Cl$_3$ and SeO$_2$ (AlfaAesar, 99.4\%), were prepared in a molar ratio of 1 : 2 : 3. Here, K$_2$CO$_3$ acts as a mineralizer which improves the quality of the single crystals (we note that any further increase of the K$_2$CO$_3$ amount in the reaction had a very little effect on size of the obtained single crystals.) For each reaction, a total of 0.5 g of reactants were mixed with 6 mL of de-ionized (DI) water. The reaction mixture was loaded to the poly(tetrafluroethylene)-lined pressure vessel and sealed tightly. After the reaction period, columnar shaped single crystals (average size 0.5 x 0.2 x 0.1 mm$^3$) were isolated using suction filtration method by washing with DI water and acetone. Each reaction produced a very homogeneous mixture of single crystals of the target phases. Phase purity and elemental composition were confirmed using powder X-ray diffraction and a Hitachi S-3400 scanning electron microscope, respectively. Due to the small size of the obtained crystals, single crystal neutron diffraction was eschewed for neutron powder diffraction. In order to obtain enough sample mass for the neutron powder diffraction experiments, these reactions were repeated multiple times to obtain 2 g of single crystals from each phase which were then gently ground to obtain powders.



\subsection{\label{subsec:scattering} Neutron Diffraction Experiments}

Neutron powder diffraction (NPD) measurements were performed using the time-of-flight high-resolution POWGEN diffractometer of Oak Ridge National Laboratory's Spallation Neutron Source \cite{Calder2018}. Using the high-intensity beam guide a wavelength band centered at 2.665 \AA\ was used to maximize flux while allowing access to the large \textit{d}-spacings needed to search for magnetic Bragg reflections.

Analysis of the neutron powder diffraction data was performed using the Rietveld method as implemented in the FullProf and GSASII software suites.\cite{Rodriguez-Carvajal1993,Toby2013} A convolution of a Pseudo-Voight and back-to-back exponentials was used in the refinements to model the instrument resolution. In addition to profile fitting, the atomic positions and isotropic atomic displacement parameters of all sites were refined. For magnetic structure determination the Simulated Annealing and Representational Analysis (SARAh) software was used as well as the Bilbao Crystallographic Server. \cite{Wills2000, Aroyo2006a,Aroyo2006b,Aroyo2011} Visualization of the nuclear crystal structure was performed using CrystalMaker\textsuperscript{\textregistered} (CrystalMaker, Software Ltd www.crystalmaker.com) while the magnetic structure was visualized using VESTA. \cite{Momma2011}

\subsection{\label{subsec:firstprinciples} First Principles Calculations}

First principles calculations were performed using the all-electron density functional theory (DFT) code 
WIEN2K \cite{Blaha2018}. The generalized gradient approximation of Perdew et al was used. \cite{Perdew1996}. To perform such calculations using an all-electron code is arduous given the very small Hydrogen muffin-tin sphere radius used here of just 0.54 Bohr (in addition to the large and low-symmetry crystal structure) and accordingly a smaller RK$_{max}$ value of just 3.12 was used. Here RK$_{max}$ is the product of the smallest sphere radius and the largest planewave expansion vector. However, despite the small RK$_{max}$ value these calculations are considered accurate as the sphere radii for O, Se and Co (and Ni) are much larger at 1.21, 1.57 and 1.92 Bohr respectively, so that the effective RK$_{max}$ for these atoms is much larger and in the range of 7.0 to over 11, in keeping with standard calculation practice.

\section{\label{sec:results} Results and Discussion}

\subsection{\label{subsec:rt} Structure}

The \textit{TM} selenite hydrate structure reported in previous structural work is shown in Fig.~\ref{fig:one}.\cite{Harrison1999,Mcmanus1991,Wildner1991,Larranaga2002} As discussed, the \textit{TM} occupies four distinct symmetry sites, one each on the special $1a: (0,0,0)$ and $1h: (\frac{1}{2},\frac{1}{2},\frac{1}{2})$ Wyckoff positions and two in the general $2i: (x,y,z)$ position. The first two positions describe a body centered like cell, with the second two defining a tilted distorted square planar-like arrangement around the $1h$ site. Each of the \textit{TM} sites is coordinated by six O atoms (with ten distinct O sites in the unit cell) making highly distorted \textit{TM}O$_6$ octahedra. In the planar arrangement of \textit{TM}, along one diagonal the octrahedra are corner sharing while along the other they are edge-sharing, this central structure is then corner-sharing with the $1a$ site octrahedra. The Se sites act as \lq linkers\rq\ between the octahedra (as shown in Fig.~\ref{fig:one}(b)) with the Se lone pair of electrons from the SeO$_3$ trigonal pyramids occupying the apparent \lq channels\rq\ running along the \textit{a} lattice direction as seen in Fig.~\ref{fig:one}(a). An insightful in-depth description of the nuclear structure is given in Ref~\onlinecite{Mcmanus1991}

\begin{figure}
	\includegraphics[width=\columnwidth]{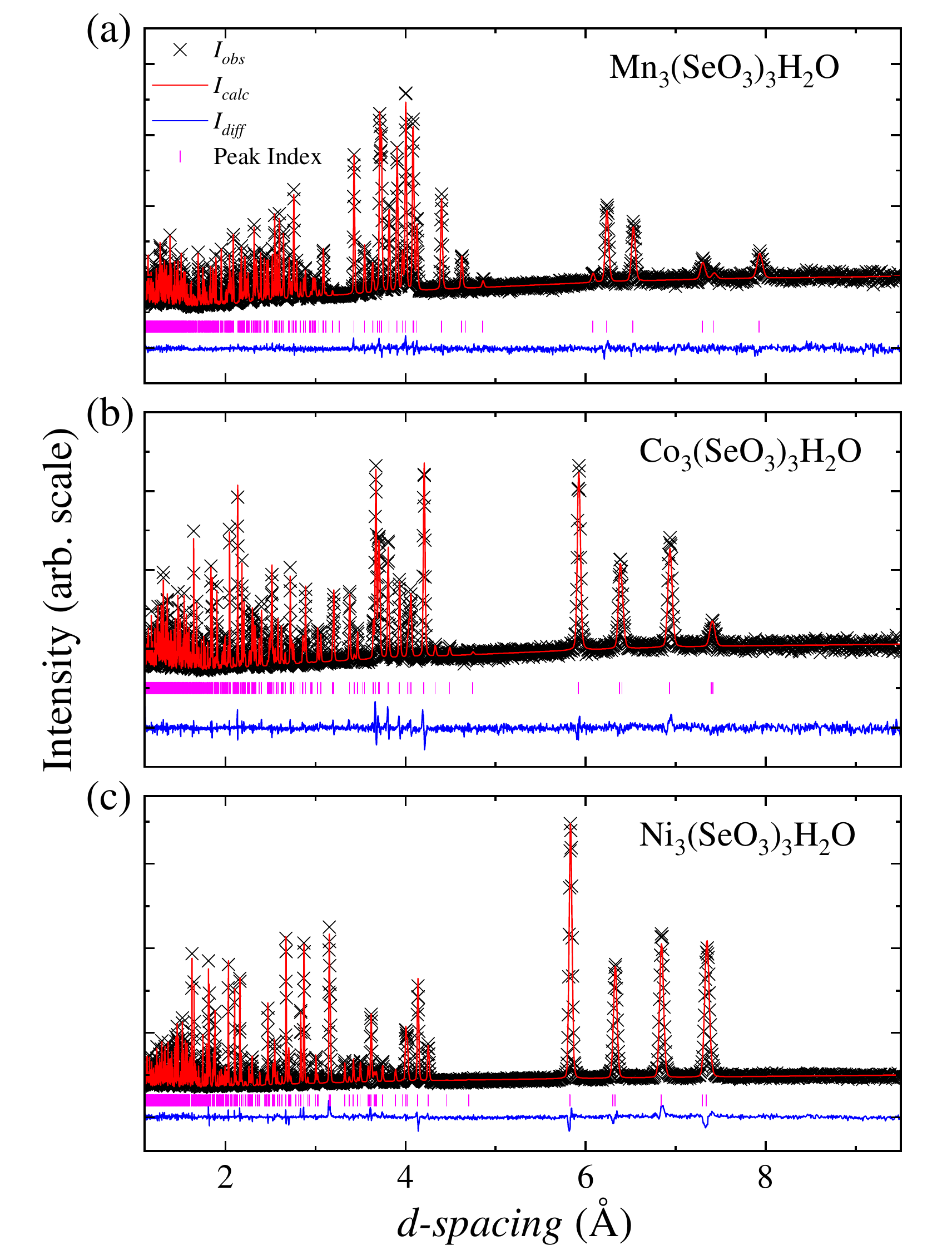}
	\caption{\label{fig:two} Best model fits using the reported triclinic crystal structure with \Pob\ symmetry to neutron powder diffraction patterns collected on POWGEN at 100, 20 and 100 K for (a) \MSOHO , (b) \CSOHO\ and (c) \NSOHO , respectively.}	
\end{figure}

Alternatively, the \textit{TM} metal sublattice can be considered as built of smaller units made of individual \textit{TM} sites, which will be helpful for later discussions involving magnetic structures in different lattice settings. \cite{Larranaga2002,Mcmanus1991} Starting with the \textit{TM}1 ($1a$) and \textit{TM}4 ($2i$) sites, a quasi-1D zigzag chain structure is seen which runs along the $(\overline{1}01)$ direction (Fig.~\ref{fig:one}(d)). These chains are then connected into planes along the \textit{bc} crystallographic plane via the \textit{TM}3 ($2i$) site. To finish the full 3D sublattice these planes are then connected via the \textit{TM}2 ($1h$) site as shown in Fig.~\ref{fig:one}(c).

\begin{table}
	\caption{\label{tab:one}Crystallographic parameters of \MSOHO , \CSOHO\ and \NSOHO\ at 100, 20 and 100 K, respectively. Parameters determined from Rietveld refinements performed using NPD collected on the POWGEN diffractometer with the frame centered at $\lambda = 2.665$ \AA . The atomic displacement parameters  are reported in units of \Asq . }
	\begin{ruledtabular}
		\begin{tabular}{llll}
     		 \multicolumn{1}{l}{\textit{TM} } & \multicolumn{1}{c}{Mn} & \multicolumn{1}{c}{Co} & \multicolumn{1}{c}{Ni} \\
	\hline
	\multicolumn{1}{l}{Space Group} & \multicolumn{1}{c}{\Pob} & \multicolumn{1}{c}{\Pob} & \multicolumn{1}{c}{\Pob} \\
	\multicolumn{1}{l}{\textit{T}} & \multicolumn{1}{c}{100 K} & \multicolumn{1}{c}{20 K} & \multicolumn{1}{c}{100 K} \\
	\multicolumn{1}{l}{$R_{wp}$} & \multicolumn{1}{c}{2.26 \%} & \multicolumn{1}{c}{2.73 \%} & \multicolumn{1}{c}{2.95 \%} \\
	\multicolumn{1}{l}{$a$ (\AA)} & \multicolumn{1}{c}{8.3257(3)} & \multicolumn{1}{c}{8.0919(5)} & \multicolumn{1}{c}{7.9823(3)} \\
	\multicolumn{1}{l}{$b$ (\AA)} & \multicolumn{1}{c}{8.2654(3)} & \multicolumn{1}{c}{8.2163(4)} & \multicolumn{1}{c}{8.1154(3)} \\
	\multicolumn{1}{l}{$c$ (\AA)} & \multicolumn{1}{c}{8.9891(4)} & \multicolumn{1}{c}{8.5401(5)} & \multicolumn{1}{c}{8.4077(3)} \\
	\multicolumn{1}{l}{$\alpha$ (deg)} & \multicolumn{1}{c}{68.714(1)} & \multicolumn{1}{c}{69.192(1)} & \multicolumn{1}{c}{69.610(1)} \\
	\multicolumn{1}{l}{$\beta$ (deg)} & \multicolumn{1}{c}{65.395(1)} & \multicolumn{1}{c}{62.820(1)} & \multicolumn{1}{c}{62.577(1)} \\
	\multicolumn{1}{l}{$\gamma$ (deg)} & \multicolumn{1}{c}{67.815(1)} & \multicolumn{1}{c}{67.217(1)} & \multicolumn{1}{c}{67.730(1)} \\
	\multicolumn{1}{l}{$V$(\AA$^3$)} & \multicolumn{1}{c}{504.99(1)} & \multicolumn{1}{c}{454.54(1)} & \multicolumn{1}{c}{437.24(1)} \\
	\multicolumn{1}{l}{\textit{TM}1 ($1a$)} 	&		&		&		\\
	\multicolumn{1}{r}{$x$}	&	0	&	0	&	0	\\
	\multicolumn{1}{r}{$y$}	&	0	&	0	&	0	\\
	\multicolumn{1}{r}{$z$}	&	0	&	0	&	0	\\
	\multicolumn{1}{r}{$U$} 	&	0.004(1)	&	0.002(2)	&	0.001(1)	\\
	\multicolumn{1}{l}{\textit{TM}2 ($1h$)} 	&		&		&		\\
	\multicolumn{1}{r}{$x$}	&	0.5	&	0.5	&	0.5	\\
	\multicolumn{1}{r}{$y$}	&	0.5	&	0.5	&	0.5	\\
	\multicolumn{1}{r}{$z$}	&	0.5	&	0.5	&	0.5	\\
	\multicolumn{1}{r}{$U$}	&	0.004(1)	&	0.002(2)	&	0.001(1)	\\
	\multicolumn{1}{l}{\textit{TM}3 ($2i$)} 	&		&      	&		\\
	\multicolumn{1}{r}{$x$}	&	0.658(1)	&	0.654(1)	&	0.654(1)	\\
	\multicolumn{1}{r}{$y$}	&	0.036(1)	&	0.036(1)	&	0.042(1)	\\
	\multicolumn{1}{r}{$z$}	&	0.795(1)	&	0.793(2)	&	0.798(1)	\\
	\multicolumn{1}{r}{$U$}	&	0.004(1)	&	0.002(2)	&	0.001(1)	\\	
	\multicolumn{1}{l}{\textit{TM}4 ($2i$)} 	&		&		&		\\
	\multicolumn{1}{r}{$x$}	&	0.695(1)	&	0.716(2)	&	0.710(1)	\\
	\multicolumn{1}{r}{$y$}	&	0.829(1)	&	0.861(2)	&	0.866(1)	\\
	\multicolumn{1}{r}{$z$}	&	0.423(1)	&	0.391(1)	&	0.390(1)	\\
	\multicolumn{1}{r}{$U$}	&	0.004(1)	&	0.002(2)	&	0.001(1)	\\
		
		\end{tabular}
	\end{ruledtabular}
\end{table}

In order to check the reported nuclear structure and see how it changes as a function of \textit{TM} we performed neutron powder diffraction measurements on \TSOHO\ with \textit{TM} = Mn, Co and Ni. NPD patterns collected at 100, 20 and 100 K for \MSOHO , \CSOHO\ and \NSOHO\ respectively together with best fit models using the reported structure are shown in Fig.~\ref{fig:two}. As seen the reported structure fits the data well, producing low residuals and visually adequate fits (Table~\ref{tab:one}). Using this structure all observed peaks are fit with no evidence of any impurity phases within the sensitivity of POWGEN. Selected crystallographic parameters extracted from the Rietveld refinements are reported in Table~\ref{tab:one}.

Despite having the same crystal structure, these three compounds have significant differences in their refined crystallographic parameters. As might be expected from considerations of the ionic radii, the progression from Mn to Co to Ni causes an overall reduction in the unit cell parameters and volume with a nearly 10\%\ volume reduction between \MSOHO\ and \CSOHO\ and further 4\%\ reduction between \CSOHO\ and \NSOHO .\cite{Shannon1976}  Looking at the individual parameters it is seen that while \textit{b}, $\alpha$ and $\gamma$ remain relatively similar across the series (at less than a 2\%\ variation between the compounds) the \textit{a}, \textit{c} and $\beta$ parameters change significantly. Changing the \textit{TM} from Mn to Ni causes a 4\%\ reduction in both \textit{a} and $\beta$ and a 7\%\ in \textit{c}. Furthermore, between \MSOHO\ and \CSOHO\ the relative lengths of the lattice parameters change with $c > a > b $ for the Mn compound while $c > b > a$ for both the Co and Ni compounds.

As a measure of these effects, one can compare the unit cell volume \textit{V} to that of a hypothetical orthogonal cell as a sort of quantification of the triclinic \lq distortion\rq\ with a ratio of 1 corresponding to no distortion. Doing so results in $V_{tri}/V_{ortho}$ of 0.82, 0.80 and 0.80  respectively for the Mn, Co and Ni compounds. By this measure, the triclinic \lq distortion\rq\ increases from \MSOHO\ to \CSOHO\ but remains fairly stable between \CSOHO\ and \NSOHO .\cite{Foadi2011}

With the complex network of corner and edge sharing \textit{TM}O$_6$ octahedra previously discussed, such anisotropic structural changes should be expected to have significant impact on the numerous potential exchange pathways between the octahedra. Tuning the anisotropy will change the relative \textit{TM}\textendash O bond lengths and \textit{TM}\textendash O\textendash \textit{TM} bond angles along different directions which may allow for a tuning of the magnetic exchange interactions. Furthermore, the anisotropic application of chemical pressure will effect the crystal field levels in the octahedra possibly changing the ground state spin of the \textit{TM}. If the structure indicates competing interactions of a similar energy scale as suggested, then small changes to the octahedra and \textit{TM}\textendash O\textendash \textit{TM} bond angles which even subtly change orbital overlap and orbital energies could have profound effects on the resulting magnetic structure.

\begin{table}
	\caption{\label{tab:two} Selected bonding parameters of the \textit{TM}4O$_6$ octahedra and along the \textit{TM}1\textendash \textit{TM}4 quasi-1D sublattice as well as average \textit{TM}\textendash \textit{TM} distances. All bond lengths and distances are reported in \AA , while bond angles are in degrees.}
	\begin{ruledtabular}
		\begin{tabular}{dddd}
     		 \multicolumn{1}{l}{\textit{TM} } & \multicolumn{1}{c}{Mn} & \multicolumn{1}{c}{Co} & \multicolumn{1}{c}{Ni} \\
	\hline
	\multicolumn{1}{l}{\textit{TM}4\textendash O }	&	&		&		\\
	\multicolumn{1}{r}{max}	&	2.40(1)	&	2.37(1)	&	2.30(1)	\\
	\multicolumn{1}{r}{min }	&	2.06(1)	&	2.00(1)	&	1.99(1)	\\
	\multicolumn{1}{l}{O\textendash \textit{TM}4\textendash O}	&		&		&		\\
	\multicolumn{1}{r}{max}	&	113.2(1)	&	109.7(1)	&	108.5(1)	\\
	\multicolumn{1}{r}{min}	&	66.24(1)	&	69.0(1)	&	74.0(1)	\\
	\multicolumn{1}{l}{\textit{TM}\textendash O\textendash \textit{TM} }	&		&		&		\\
	\multicolumn{1}{r}{1\textendash 4}	&	113.5(1)	&	95.78(1)	&	95.34(1)	\\
	\multicolumn{1}{r}{4\textendash 4}	&	95.5(1)	&	96.84(1)	&	96.73(1)	\\
	\multicolumn{1}{l}{\textit{TM}\textendash \textit{TM}}	&		&		&		\\
	\multicolumn{1}{r}{avg}	&	3.74(1)	&	3.62(1)	&	3.54(1)	\\
		
		\end{tabular}
	\end{ruledtabular}
\end{table}

Therefore, despite the inherent difficulty in quantifying such changes in this low symmetry structure, it is worth considering single elements to see the scale and trend of these effects. To start, we consider a single octahedron and measure the distortion via the maximum and minimum \textit{TM}-O bond lengths and \textit{TM}\textendash O\textendash \textit{TM} bond angles. As suggested in ref.~\onlinecite{Mcmanus1991} the \textit{TM}4 octahedra is the most distorted for all of the measured compounds and so we will focus there. As shown in Table~\ref{tab:two}, there are significant changes both in the raw values between the three systems and in the relative differences between the maximal bond measures. In general, one sees a trend towards a less distorted octahedra going from Mn to Co to Ni. In particular, the maximal O\textendash \textit{TM}\textendash O angles become significantly less dissimilar while the \textit{TM}\textendash O bond lengths stay relatively stable. If we similarly consider the \textit{TM}1\textendash \textit{TM}4 quasi-1D chain structure, it is seen that the \textit{TM}\textendash O\textendash \textit{TM} bond angles responsible for any superexchange interaction shift significantly - most notably in the relative value of the two possible exchange paths (\textit{TM}1\textendash O3\textendash \textit{TM}4 and \textit{TM}4\textendash O4\textendash \textit{TM}4) which flip in relative magnitude between the Mn compound and the Co/Ni compounds. As will be discussed in general terms in the DFT analysis, such structural tunability in a system with expected competing magnetic interactions may give a way to realize different magnetic structures via different \textit{TM} and allow for tuning between them via doping continuously between the respective \textit{TM} compounds. \cite{Geertsma1990}

\subsection{\label{subsec:mag} Zero-Field Magnetic Structures}

Aside from some preliminary magnetic susceptibility measurements on \MSOHO ,  very little work has been performed to actually characterize the magnetic properties of these compounds. \cite{Choudhury2002, Kovrugin2018, Larranaga2002} As we show here, these materials have a rich magnetic behavior with indications of moderate frustration and a tunability via \textit{TM} selection.

\begin{figure}
	\includegraphics[width=\columnwidth]{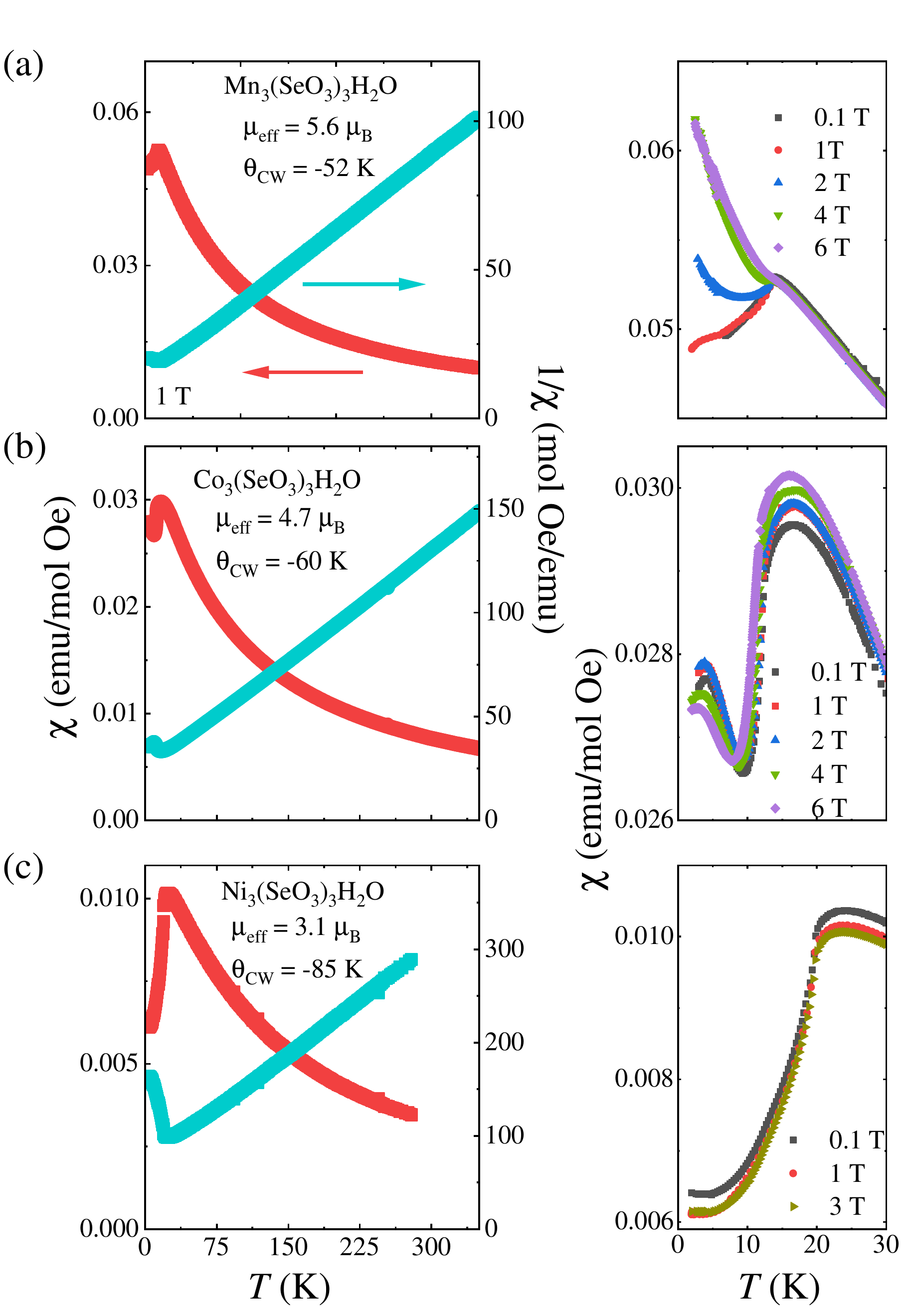}
	\caption{\label{fig:three}Magnetic susceptibility $\chi$ and the inverse susceptibility ($1/\chi$) for (a) \MSOHO , (b) \CSOHO\ and (c) \NSOHO  . The scale for $\chi$ is shown on the left axis while the scale for $1/\chi$ is shown on the right axis. All samples where measured under an applied field of 1 T. Plotted in the right panels are the low temperature data together with measurements taken under different applied fields.}	
\end{figure}

In Fig.~\ref{fig:three} we show the results of magnetic susceptibility measurements performed between 350 and 2 K for \MSOHO , \CSOHO\ and \NSOHO\ along with their inverse susceptibility. The high temperature range of the inverse susceptibility ($1/\chi(T > 100 \text{ K}))$ appears linear allowing the use of Curie-Weiss law fitting to characterize the magnetism of these compounds. Using such fitting, we report the effective moment per \textit{TM} ($\mu_B/TM$) and Curie-Weiss temperature ($\theta_{CW}$) for each compound in Fig.~\ref{fig:three}.

Starting with \MSOHO\ (Fig.~\ref{fig:three}(a)), we find the magnetic susceptibility to increase with decreasing temperature until $\sim 15$ K where there is a sudden downturn likely indicating the onset of AFM order (with $T_N \sim\ 15$ K). From the Curie-Weiss Law fit, we find $\theta_{CW}= -52$ K which both indicates AFM interactions between the magnetic sites and moderate magnetic frustration (with a frustration index $f = |\theta_{CW}|/T_N$ of 3.5).  These results agree reasonably well with those reported in Ref.~\onlinecite{Larranaga2002}. From our fitting we find an effective moment of 5.6 $\mu_B$/Mn which is consistent with a Mn$^{2+}$ ion in a high spin state.

For \NSOHO\ we see similar behavior with $\chi$ increasing until $\sim 21$ K and a $\theta_{CW}$ of -85 K. This again indicates AFM interactions/order and a moderate frustration index of 3.8 similar to \MSOHO\ . The Curie-Weiss fit suggests an effective moment per Ni of $3.1 \mu_B$ which is close to that expected for Ni$^{2+}$.

\CSOHO\ shows similar high temperature behavior with a sudden downturn at $\sim 20$ K, $\theta_{CW}$ indicative of AFM interactions, frustration index of 3 and an effective moment consistent with the Co$^{2+}$ high spin state. However, below the apparent AFM transition, the susceptibility does not exhibit the monotonic decrease with temperature seen in the other compounds. Instead, $\chi_{Co}$ oscillates reaching a local minimum at $~10$ K then again increasing with decreasing temperature before starting to decrease with temperature at the lowest measured temperatures. These multiple local extrema may indicate several competing ground states and multiple magnetic transitions.

These results encourage neutron diffraction experiments to more fully characterize the potential magnetic orders as well as learn how they change between the three compounds. Therefore, we collected NPD patterns on all three samples between 100 and 2 K to look for magnetic Bragg peaks expected from the susceptibility measurements and solve any such magnetic structure.

\begin{figure}
	\includegraphics[width=\columnwidth]{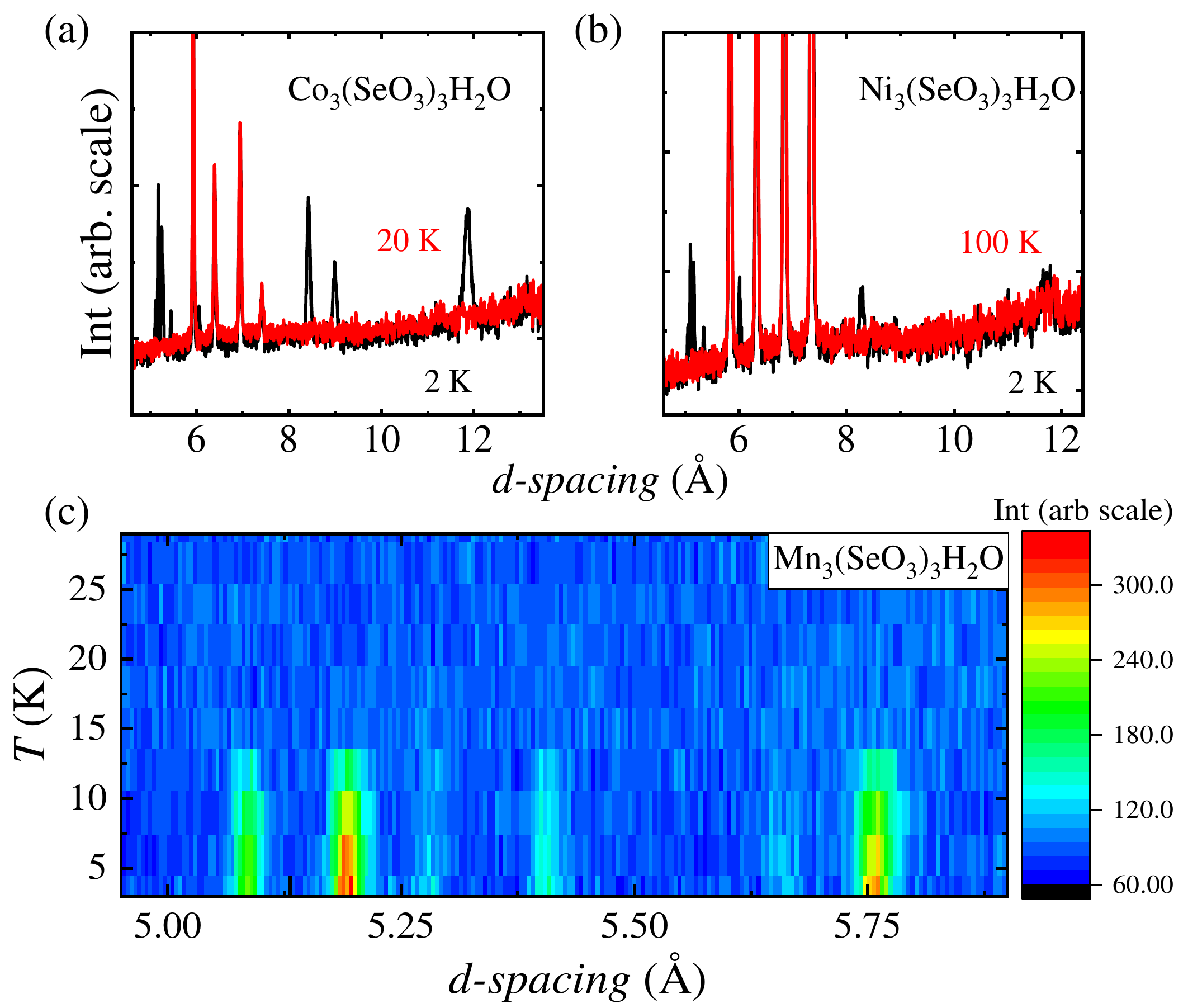}
	\caption{\label{fig:four} Comparisons of the high \textit{d}-spacing regions of NPD patterns collected in the paramagnetic state and below the putative magnetic transition for (a) \CSOHO\ and (b) \NSOHO . (c) A NPD diffractogram for \MSOHO\ collected on warming from base temperature (2 K) to above the cusp feature seen in the magnetic susceptibility.}	
\end{figure}

Fig.~\ref{fig:four} compares NPD patterns collected above and below the features seen in the susceptibility measurements, focusing on the high \textit{d}-spacing range where magnetic scattering is generally strongest. In all three samples new Bragg reflections are seen upon cooling below the susceptibility down turn. For the \MSOHO\ sample, the large size of the Mn$^{2+}$ moment allowed for quicker collection times and so the new Bragg peaks could be followed pseudo-continuously through the supposed $T_N$, the result is plotted as a diffractogram in Fig.~\ref{fig:four}(c). As seen the appearance of the new reflections coincide with the temperature of the cusp in the susceptibility ($\sim 15$ K), indicating that this feature is indeed a magnetic transition. As new peaks appear at \textit{d}-spacings not allowed by the nuclear structure we can characterize this as an AFM transition as suggested by the downturn seen in the susceptibility data. We note that no additional features are seen in the NPD patterns of \CSOHO , which correspond to the lower temperature non-monotonic behavior observed in the susceptibility. As will be confirmed later, we take this as indicating only a single magnetic transition in this compound.

\begin{table}
	\caption{\label{tab:three} Irreducible representations ($\Gamma$), Mn sites active for the $\Gamma$, number of basis vectors per \textit{TM} site ($\psi$) and magnetic space groups for the \Pob\ nuclear symmetry with $k = (0,\frac{1}{2},\frac{1}{2})$ or $(\frac{1}{2},\frac{1}{2},0)$.}
	\begin{ruledtabular}
		\begin{tabular}{cccc}
     		 \multicolumn{1}{c}{$\Gamma$} & Mn site & \multicolumn{1}{c}{$\psi$} & \multicolumn{1}{c}{Magnetic space group} \\
	\hline
	\multirow[t]{4}{*}{$\Gamma_1\ $} & 1  & 3 & $P\overline{1}$     \\
									 & 2  & 3 &          \\
									 & 3  & 3 &          \\
									 & 4  & 3 &          \\
	\multirow[t]{4}{*}{$\Gamma_2\ $} & 3  & 3 & $P\overline{1}'$     \\
									 & 4  & 3 &         \\

		\end{tabular}
	\end{ruledtabular}
\end{table}

In order to further characterize the magnetic structure, we turn to identifying a magnetic ordering vector ($k$). Considering the \textit{d}-spacings of the new peaks and the nuclear unit cell it is possible to index all the new low temperature peaks by applying factors of two to the nuclear unit cell. Doing so we find two $k$ vectors with $k =(0, \frac{1}{2}, \frac{1}{2})$ for \MSOHO , and $k = (\frac{1}{2}, \frac{1}{2}, 0)$ for both \CSOHO\ and \NSOHO . It is notable that these three materials are bipartite and do not all realize a single $k$-vector. We tentatively note that this correlates to several previously discussed structural differences which grouped these materials similarly such as the relative lengths of the lattice parameters with $a > b$ for the Mn compound but $b > a$ for both Co and Ni.

Shown in Table~\ref{tab:three} are the results of representational analysis using the determined $k$ vectors, the \Pob\ nuclear symmetry and the positions of the four \textit{TM} ions for $k =(0, \frac{1}{2}, \frac{1}{2})$ and $k = (\frac{1}{2}, \frac{1}{2}, 0)$. We note that representational analysis using either $k$ and the \Pob\ symmetry results in identical constituent irreducible representations and so we only show a single table. The difference then between the resulting structures is found in the direction of the AFM correlations as determined by the distinct $k$-vectors. For each $k$ we find only two possible irreducible representations ($\Gamma$): $\Gamma_1$ and $\Gamma_2$. Each $\Gamma$ has three basis vectors ($\psi$) per \textit{TM} site, with each basis vector corresponding to a magnetic moment component along one of the crystallographic axes. For $\Gamma_1$ this essentially allows each site to have a freely refinable magnetic moment which can have components along all three of the crystal axes with the \textit{k}-vector describing the nature (AFM or FM) of the intrasite correlations.  On the other hand, the $\Gamma_2$ irrep is only active for the \textit{TM}3 and \textit{TM}4 sites, enforcing a magnetic moment of zero on the \textit{TM}1 and \textit{TM}2 sites by symmetry. We find that this model produces poor fits of the data in addition to being physically unlikely, and so focus only on the $\Gamma_1$ structure.

\begin{figure}
	\includegraphics[width=\columnwidth]{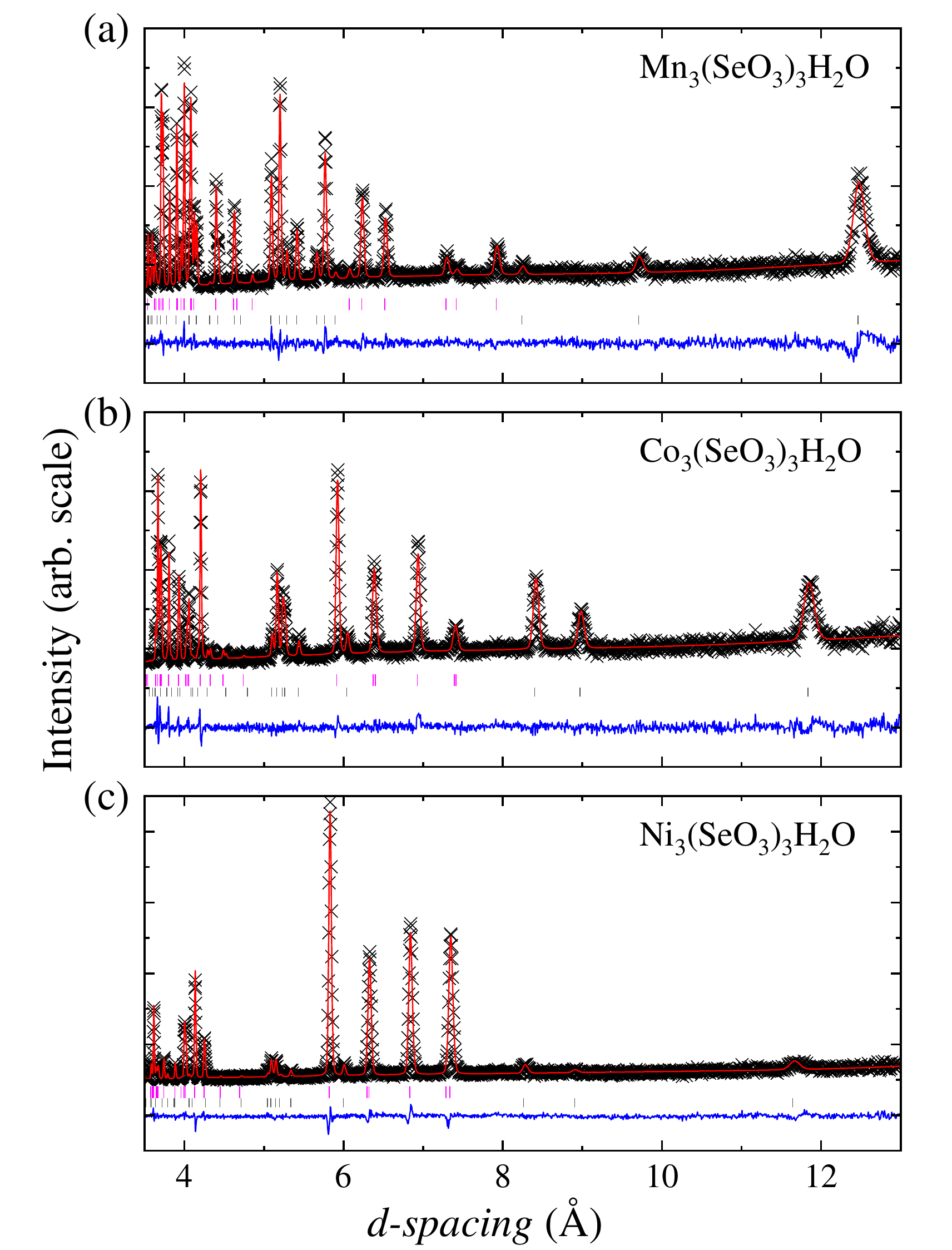}
	\caption{\label{fig:five} Best model fits using the reported triclinic nuclear structure and the $\Gamma_1$ magnetic structures to neutron powder diffraction patterns collected on POWGEN at 2 K for (a) \MSOHO , (b) \CSOHO\ and (c) \NSOHO\ respectively. Peaks belonging to the nuclear and magnetic structures are indicated by magenta and black tick marks respectively. As in Fig.~\ref{fig:five}, the black \lq x\rq\ markers, red lines and blue lines represent the data, model and difference curves respectively.}	
\end{figure}



Shown in Fig.~\ref{fig:five} are the NPD data collected at 2 K for each compound together with the calculated diffraction pattern for the best fit model determined from Rietveld refinements using the $\Gamma_1$ irrep. As seen (and reported quantitatively in Table~\ref{tab:four}), the models using $\Gamma_1$ are capable of reproducing well the positions and intensities of all the new low temperature reflections for all three compounds.  Comparing the crystallographic parameters extracted from the low temperature fits we generally find little change between the PM and AFM states in either the lattice parameters or the \textit{TM} sites' fractional coordinates. Considering the low temperatures of the PM data this is not surprising as little thermal expansion should be expected below 100 K. It does suggest however, that either there is little magnetoelastic coupling or, as our DFT results will suggest, that magnetic frustration may already be affecting the nuclear structure even above \Tn . 

Looking at the extracted magnetic moment magnitude, we see a somewhat significant reduction from the effective moments found in the Curie-Weiss fitting of the magnetic susceptibility in the PM state. The magnetic moments determined from the Rietveld refinements are $\sim  75, 57$ and $74 \%$ the values found from the Curie-Weiss fitting for \MSOHO , \CSOHO\ and \NSOHO , respectively. A similar reduction of the ordered moment from the expected free ion moment has been seen in the rare-earth pyrochlores where it was a sign of magnetic frustration. \cite{Dun2015,Li2014} However, if we instead compare the refined moment  to the calculated moment for \textit{TM} with $d^3, d^5 \text{ and } d^8$  (i.e. Mn$^{2+}$, Co$^{2+}$ and Ni$^{2+}$) with quenched orbital angular momentum ($M=g S \mu_B$) we find differences of $< 15\%$. Considering the differences between the effective and refined moments, it is interesting that \CSOHO\ has the largest reduction in the observed magnetic moment in the ordered state. This compound also exhibits a non-monotonic behavior in its low temperature susceptibility - these observations together may indicate competing grounds states leading to strong frustration. However, further work fully explicating the spin-Hamiltonian is needed to elucidate potential competing interactions in these materials. 

We note that while we do report variations between the moments on the four distinct \textit{TM} sites in each compound which could ostensibly indicate different valences or spin states due to different crystal field levels seen by different ions, the error on our reported moments is relatively large and prevents us from ascribing a strong certainty to this variation. However, as will be discussed later, our DFT results suggest that this is certainly a possibility, especially in the Co compound. More work using single crystals for neutron diffraction measurements could help reduce these uncertainties and determine whether the distinct sites have different spin states. 

\begin{table}
	\caption{\label{tab:four}Crystallographic parameters of the magnetic phases of \MSOHO , \CSOHO\ and \NSOHO\ at 2 K. Parameters determined from Rietveld refinements performed using NPD collected on the POWGEN diffractometer with the frame centered at $\lambda = 2.665$ \AA . The atomic displacement parameters  are reported in units of \Asq\ and magnetic moments $M$ are reported in $\mu_B/TM$. The unit cell and atomic positions are reported in the nuclear unit cell setting for ease of comparison. The magnetic space group \Pob\ is listed in the Belov-Neronova-Smirnova notation and corresponds to magnetic space group $\# 2.4$.}
	\begin{ruledtabular}
		\begin{tabular}{llll}
     		 \multicolumn{1}{l}{\textit{TM} } & \multicolumn{1}{c}{Mn} & \multicolumn{1}{c}{Co} & \multicolumn{1}{c}{Ni} \\
	\hline
	\multicolumn{1}{l}{Mag. Space Group} & \multicolumn{1}{c}{\Pob} & \multicolumn{1}{c}{\Pob} & \multicolumn{1}{c}{\Pob} \\
	\multicolumn{1}{l}{\textit{T}} & \multicolumn{1}{c}{2 K} & \multicolumn{1}{c}{2 K} & \multicolumn{1}{c}{2 K} \\
	\multicolumn{1}{l}{$R_{wp}$} & \multicolumn{1}{c}{2.5 \%} & \multicolumn{1}{c}{2.7 \%} & \multicolumn{1}{c}{3.2 \%} \\
	\multicolumn{1}{l}{$a$ (\AA)} & \multicolumn{1}{c}{8.3219(3)} & \multicolumn{1}{c}{8.0913(5)} & \multicolumn{1}{c}{7.9799(3)} \\
	\multicolumn{1}{l}{$b$ (\AA)} & \multicolumn{1}{c}{8.2636(3)} & \multicolumn{1}{c}{8.2165(4)} & \multicolumn{1}{c}{8.1153(3)} \\
	\multicolumn{1}{l}{$c$ (\AA)} & \multicolumn{1}{c}{8.9881(4)} & \multicolumn{1}{c}{8.5402(5)} & \multicolumn{1}{c}{8.4048(3)} \\
	\multicolumn{1}{l}{$\alpha$ (deg)} & \multicolumn{1}{c}{68.670(1)} & \multicolumn{1}{c}{69.191(1)} & \multicolumn{1}{c}{69.610(1)} \\
	\multicolumn{1}{l}{$\beta$ (deg)} & \multicolumn{1}{c}{65.379(1)} & \multicolumn{1}{c}{62.823(1)} & \multicolumn{1}{c}{62.570(1)} \\
	\multicolumn{1}{l}{$\gamma$ (deg)} & \multicolumn{1}{c}{67.830(1)} & \multicolumn{1}{c}{67.215(1)} & \multicolumn{1}{c}{67.725(1)} \\
	\multicolumn{1}{l}{$V$(\AA$^3$)} & \multicolumn{1}{c}{504.47(1)} & \multicolumn{1}{c}{454.53(1)} & \multicolumn{1}{c}{436.93(1)} \\
	\multicolumn{1}{l}{\textit{TM}1 ($1a$)} 	&		&		&		\\
	\multicolumn{1}{r}{$x$}	&	0	&	0	&	0	\\
	\multicolumn{1}{r}{$y$}	&	0	&	0	&	0	\\
	\multicolumn{1}{r}{$z$}	&	0	&	0	&	0	\\
	\multicolumn{1}{r}{$U$} 	&	0.003(2)	&	0.003(2)	&	0.002(2)	\\
	\multicolumn{1}{r}{$M$} 	&	4.2(2)	&	3.0(3)	&	2.3(5)	\\
	\multicolumn{1}{l}{\textit{TM}2 ($1h$)} 	&		&		&		\\
	\multicolumn{1}{r}{$x$}	&	0.5	&	0.5	&	0.5	\\
	\multicolumn{1}{r}{$y$}	&	0.5	&	0.5	&	0.5	\\
	\multicolumn{1}{r}{$z$}	&	0.5	&	0.5	&	0.5	\\
	\multicolumn{1}{r}{$U$}	&	0.003(2)	&	0.003(2)	&	0.002(2)	\\
	\multicolumn{1}{r}{$M$} 	&	4.3(2)	&	2.3(4)	&	2.0(5)	\\
	\multicolumn{1}{l}{\textit{TM}3 ($2i$)} 	&		&      	&		\\
	\multicolumn{1}{r}{$x$}	&	0.658(2)	&	0.651(1)	&	0.654(1)	\\
	\multicolumn{1}{r}{$y$}	&	0.035(1)	&	0.033(1)	&	0.039(1)	\\
	\multicolumn{1}{r}{$z$}	&	0.795(2)	&	0.793(2)	&	0.800(1)	\\
	\multicolumn{1}{r}{$U$}	&	0.003(2)	&	0.003(2)	&	0.002(2)	\\	
	\multicolumn{1}{r}{$M$} 	&	3.9(2)	&	2.7(3)	&	2.9(4)	\\
	\multicolumn{1}{l}{\textit{TM}4 ($2i$)} 	&		&		&		\\
	\multicolumn{1}{r}{$x$}	&	0.698(2)	&	0.718(2)	&	0.709(1)	\\
	\multicolumn{1}{r}{$y$}	&	0.826(2)	&	0.863(2)	&	0.867(1)	\\
	\multicolumn{1}{r}{$z$}	&	0.422(2)	&	0.389(2)	&	0.389(1)	\\
	\multicolumn{1}{r}{$U$}	&	0.003(2)	&	0.003(2)	&	0.002(2)	\\
	\multicolumn{1}{r}{$M$} 	&	4.5(2)	&	2.8(3)	&	1.9(5)	\\
		
		\end{tabular}
	\end{ruledtabular}
\end{table}

\begin{figure*}
	\includegraphics[width=\textwidth]{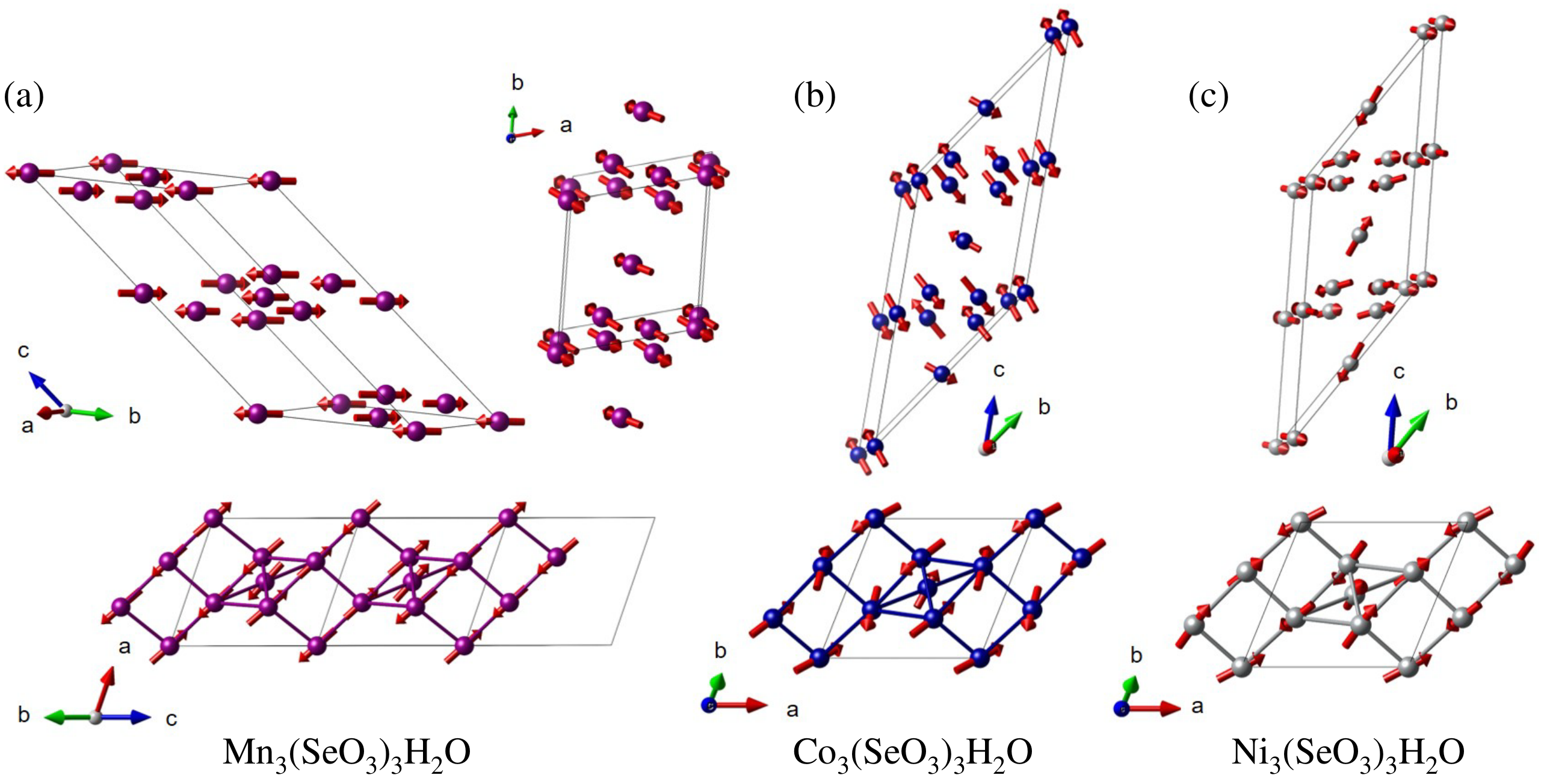}
	\caption{\label{fig:six} Magnetic structures determined from the best fit models found in Rietveld refinments using the POWGEN NPDs collected at 2 K for (a) \MSOHO , (b) \CSOHO\ and (c) \NSOHO . All structures are plotted in their magnetic unit cells which are different from the nuclear unit cell due to the non-zero \textit{k} vectors. To ease in comparison, all structures have been plotted along the equivalent crysatllographic directions to those shown in Fig.~\ref{fig:one}. For \MSOHO , an additional viewing direction was added to discuss the structures unique co-planer behavior.}	
\end{figure*}

The magnetic models obtained from the best fit Rietveld refinements are shown in Fig.~\ref{fig:six}. Here they have been plotted using the expanded magnetic unit cells (transformed unit cells of $(-c,-a+b,2b)$ for \MSOHO\ and $(a,b-c,2c)$ for \CSOHO\ and \NSOHO ). To allow for ease of comparison despite the different ordering vectors the unit cells have been orientated along directions, which allow all three compounds to be compared to the figures of the nuclear unit cell shown in Fig.~\ref{fig:one}(c) and (d).

\MSOHO\ has the simplest magnetic structure, only requiring two of the three possible $\psi$ per \textit{TM} leading to the co-linear order shown in Fig.~\ref{fig:six}(a). Here the Mn atoms form corrugated planes at the $z = 0, \frac{1}{2}$ positions which are parallel to the magnetic cell's \textit{ab} plane and stack with AFM correlations. However, since this magnetic unit cell is unique to \MSOHO\ and does not compare readily to the Co and Ni compounds. Therefore we also show \MSOHO 's magnetic structure along the crystallographic directions established in Fig~\ref{fig:one} in which it is easier to discern the previously described \textit{TM} sublattices. In this view, we find AFM correlations along the Mn1-Mn4 chain structure. This chain is then linked AFM to the neighboring Mn1-Mn4 chain through the Mn3 which is FM to the Mn4 site and AFM to the neighboring chain's Mn1. This Mn1-Mn4-Mn3 structure creates the planes which are then linked through the Mn2 site. Using this description, we find the layers to be FM correlated along the stacking direction.

For \CSOHO\ and \NSOHO\ no simple picture of the magnetic structure emerges as it does for \MSOHO . In both compounds each symmetry independent \textit{TM} site requires non-zero coefficients on all three available $\psi$  to produce adequate fits of the data. The resulting structures are neither purely co-linear nor co-planer and are most easily described via the quasi-1D and quasi-2D sublattices defined earlier.

Starting with the quasi-1D chains of \CSOHO , we find FM correlations along the Co1-Co4 links and AFM correlations along the Co4-Co4 link with the moments pointing approximately perpendicular to the Co4-Co4 direction. This creates chain segments of three FM aligned Co4-Co1-Co4 coupled AFM to each other through a Co4-Co4 link (note that because of the reorientation of the magnetic unit cell the Co1 site at the bottom right corner of Fig.~\ref{fig:six}(b) is not symmetry equivalent to the Co1 at the top left corner - it only appears that way do to projecting along the \textit{c} axis, this is clear in the perpendicular view.) The Co1-Co4 chains are then linked as before through nominally FM correlations along the short Co4-Co3 direction creating the Co1-Co4-Co3 planes. Within these planes the Co moments are nearly co-linear and canted out of the plane. The Co2 sites which link these planes are less canted from this stacking direction and are AFM correlated from plane to plane.

For \NSOHO , the interactions along the Ni4-Ni1-Ni3 planes and between planes can generally be described similarly to \CSOHO . This is most easily seen in the projection along the magnetic unit cell's \textit{c}-axis (Fig.~\ref{fig:six}(c)) where we see similar FM Ni4-Ni1-Ni4 segments AFM aligned through a Ni4-Ni4 link and FM connected to each other through the short Ni4-Ni3. However, as seen in the perpendicular view, the Ni1-Ni4-Ni3 moments show much less canting than in \CSOHO\ and are nearly co-linear along a chain made of the Ni1-Ni3-Ni4 sites.

As described, there are significant differences in the magnetic structures realized in these three compounds. Starting with \MSOHO\ there is a clear easy-axis type behavior where all four Mn sites align either up or down along a globally determined direction (the crystallographic $(1\overline{1}0)$ direction in the magnetic unit cell). In the Co compound this is clearly not the case as the Co2 and Co1-Co4-Co3 planes have different moment directions. Even in the Co1-Co4-Co3 plane we find no single easy-axis, with the Co1 and Co4 sites appearing co-linear while the Co3 site is rotated with respect to the rest of the plane. The easy-axis description also breaks down for the Ni compound where no two sites share a moment direction. While in the Ni1-Ni4-Ni3 plane the moments' loosely point along the Ni1-Ni3 and Ni-Ni4 directions, the Ni2 site is rotated significantly out of this direction.

Correlating these changes to the previously discussed structural changes is quite difficult in such a low symmetry structure. Previously, we defined a potential measure of the triclinic distortion as a volume ratio and found that \MSOHO\ had the smallest such distortion while \CSOHO\ and \NSOHO\ were more distorted and similar to one another. Considering the magnetic structure, we see that the least distorted structure (by this measure) has the most regular magnetic order. On the other hand, the bonding parameters of the \textit{TM} octahedra were found to be the most highly distorted in \MSOHO\ when compared to \CSOHO\ and \NSOHO\ and so such simple considerations are not too illuminating here (Table~\ref{tab:two}). Similarly, if we try to apply the Goodenough-Kanamori rules based off the \textit{TM}\textendash O\textendash\textit{TM} bond angles reported in Table~\ref{tab:two} we generally do not find good agreement finding for instance AFM correlations along bond angles close to 90\degrees. Looking at the average \textit{TM}-\textit{TM} distance we find that the magnetic structure becomes less co-linear as the distance decreases. This may be indicate an increase in orbital overlap in the Ni and Co compounds, which could strengthen both direct and superexchange interactions relative to single ion physics and cause the non co-linear structures we observe. This would suggest that single ion anisotropies may ostensibly be stronger in the Mn compound and help establish the easy-axis behavior. However, a more complete understanding of what leads to these changes would require careful inelastic neutron scattering experiments to determine both the exchange interactions and local anisotropies of the magnetic Hamiltonian - we leave this to future neutron scattering work.

\subsection{\label{subsec:DFT} First Principles Analysis of the Magnetic Structures}

In an effort to shed light on the observed magnetic behavior, we performed DFT calculations studying the Co and Ni compounds. For the crystal parameters we used the experimentally determined lattice parameters and angles, but relaxed the internal parameters in a non-magnetic configuration.  This produces internal bonding parameters somewhat different than those determined experimentally, in terms of \textit{TM}-O distances. This suggests either the influence of magnetoelastic coupling in determining these structures, or potentially the difficulty in applying standard DFT approaches to these compounds. Note that recent work \cite{Chen2019,Pokharel2018} has found that for magnetically frustrated materials as considered here, the magnetic order can play an important role in determining the crystal structure, even well above the ordering temperature, and it is quite possible that similar effects are relevant here. The results presented here are to be considered in this context. 
 
For both compounds,  FM and AFM configurations were considered with all magnetic states falling more than 200 meV or 450 meV per 3\textit{d} atom (for the Co and Ni compounds respectively) below the non-magnetic configurations. Therefore, despite the relatively low ordering temperatures, the magnetism can be considered as local moment magnetism. For both compounds the AFM states were lower in energy than the FM states in agreement with the experimentally determined AFM structures. However, it is important to note that we only considered co-linear AFM states in our calculations due to the computational challenge of handing these low symmetry nuclear structures with canted magnetism.  

Starting with \CSOHO , we find the magnitude of the ordered moments for the Co1, Co2 and Co3 sites in the ground-state AFM configuration are 2.45, 2.31 and 2.45 $\mu_B$ respectively in reasonable agreement with the refined values. On the other hand, we find a moment on the Co4 site of 0.28 $\mu_B$ - significantly less than the 2.8 $\mu_B$ found experimentally. This is surprising, while it is possible that we observe a reduction in one of the Co sites experimentally, it is the Co2 site not the Co4 site and the effect, if real, is significantly less than seen here. It is likely that this reduction originates from the substantial structural differences among the various Co sites which will produce different local environments and possibly spin-states. It is possible that the internal parameter relaxation applied in our calculations gives rise to this discrepancy with experiment. However an important take away is then that the local bonding can lead to such a change in the local moment on ostensibly isovalent sites and that the employed model is likely missing some interaction (such as magnetoelastic coupling) as suggested previously and seen in other similar systems. \cite{Chen2019, Pokharel2018} We also note, that in our calculations we found a significant reduction in energy when the Co3 and (moreso) Co4 $2i$ sites were allowed to split. This led to differing moments on Co3/Co3' and Co4/Co4' sites. However, since this is not observed experimentally we compare to a single value on either of these two sites.

For \NSOHO\ we also find an AFM ground state. Here the magnitude of the ordered moments on the Ni1-4 sites, respectively, are 1.53, 1.54, 1.54 and 1.48 $\mu_{B}$. While these are smaller than the observed values, which range from 1.9 to 2.9 $\mu_B$, we note that the calculated values are understated due to the small muffin tin radius. A fair estimate of the full ordered moment can be made from the FM calculation, which finds a total moment of 12 $\mu_B$ per cell, or 2 $\mu_B$, which is significantly closer to the 1.9 to 2.9 $\mu_B$.  This presents a substantial difference from the \CSOHO\ results, where the Co4 moment (in the ferromagnetic configuration) is just 0.38 $\mu_B$.

The implication of these results is that, as found in neutron diffraction, there are substantial differences in the magnetic order of the two compounds, despite a nominally similar physical structure and that small changes to the internal bonding parameters can have vast effects on the magnetic behavior. In the calculated ground states both compounds are insulating, so that simple charge counting would yield a divalent state for both 3\textit{d} atoms, with the extra paired 3\textit{d} electron of Ni accounting for the reduced moment in the Ni compound. This extra electron of the Ni element clearly drives the substantial differences in magnetic order between the two compounds, with the magnetic order already complex due to the low-symmetry crystal structure in a quaternary oxide.

\subsection{\label{subsec:mag} Field-Dependent Magnetism}

\begin{figure}
	\includegraphics[width=\columnwidth]{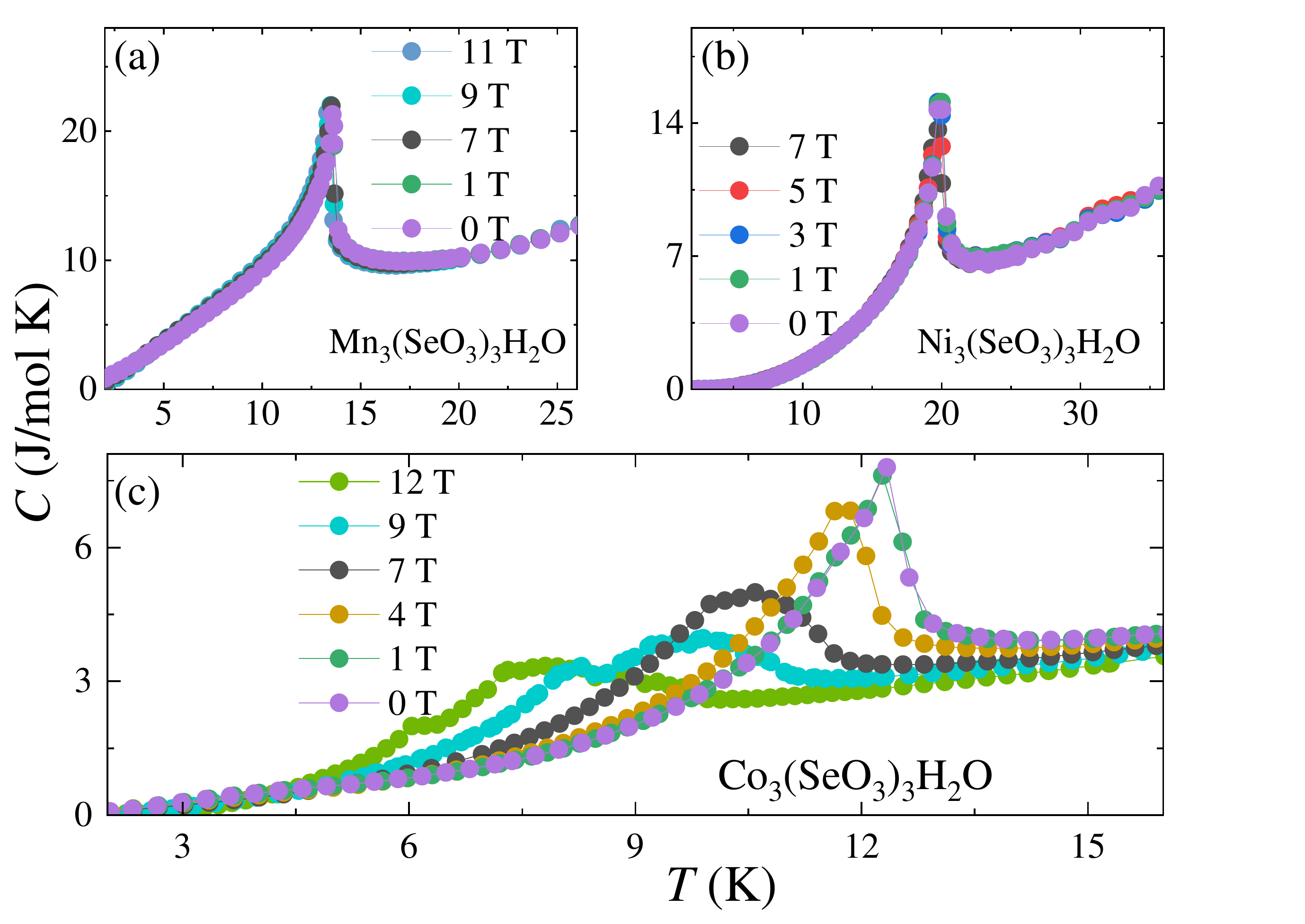}
	\caption{\label{fig:seven} Heat capacity collected under applied magnetic fields on pelletized powder samples of (a) \MSOHO , (b) \NSOHO\ and (c) \CSOHO .}	
\end{figure}

To better characterize the strength of the magnetic interactions, we performed field dependent magnetic susceptibility and heat capacity measurements (Figs.~\ref{fig:three} and \ref{fig:seven}). Such measurements may allow us to identify the relative scales of magnetic interactions via observing metamagnetic transitions and the field strengths needed to influence the magnetic order

For \MSOHO\ we find a metamagnetic transition in the magnetic susceptibility which occurs between 2 and 4 T. Here we see the temperature of the susceptibility feature does not change significantly, but the behavior below $T_N$ goes from a down turn indicating AFM order to an upturn indicating FM type order. Looking at the heat capacity (Fig.~\ref{fig:seven}(a)) we can be sure this is not a change to PM as the sharp cusp feature observed at zero field persists up to the highest measured field of 11 T. While we cannot be sure from our current measurements, an explanation consistent with these data is that the Mn sites' main interaction is single-ion anisotropies which determine locally the moment direction. As the field is increased these moments may undergo a spin flip transition (as we recently reported for a local-Ising type system (\onlinecite{Taddei2019})), which would allow them to align somewhat with the field while remaining distinct from a polarized PM state as indicated by the sharp feature in the heat capacity.

On the other hand, \CSOHO\ shows relatively little change in its susceptibility under fields up to 6 T (Fig.~\ref{fig:three}(b)). No upturn indicating FM like behavior is observed. Instead, the peak indicative of $T_N$ and the previously discussed lower temperature features are slightly suppressed to lower temperatures with increasing field. However, the heat capacity tells a different story. At zero field a sharp transition is observed (as for \MSOHO ) with no evidence of the second transition tentatively identified in the susceptibility data, confirming our analysis of the NPD data (Fig.~\ref{fig:seven}(c)). As a magnetic field is applied and increased this peak broadens and decreases in temperature, indicating a suppression of $T_N$ with field. However, by 9 T this heat capacity feature has broadened considerably and appears to have split into two peaks. Upon increasing the field further (to 12 T), this behavior continues with the both peaks being pushed to lower temperatures and the two peak feature persisting. 

Such behavior is difficult to interpret in light of the susceptibility data. With no field dependent upturn to the susceptibility we can suppose that the field is not driving a spin-flip transition (even of some part of the Co sublattice.) However, the two peak feature indicates the applied field must be inducing some second transition, whether it is due to the random field orientation with respect to individual crystallites in the measured powder or is intrinsic from the whole sample undergoing multiple consecutive transitions is difficult to say. However, due to the already low crystal symmetry (\Pob ) and low temperatures involved we argue that this is likely an induced magnetic transition rather than a structural one. Therefore, we suggest that such results indicate the presence of multiple AFM states accessible with relatively low fields and so merit future neutron diffraction work with single crystal samples.

Finally, the field dependent measurements of \NSOHO\ indicate behavior different than either of these samples. In the susceptibility measurements nearly no change in the sample response is found with neither the FM behavior of \MSOHO\ mor the suppression of $T_N$ seen in \CSOHO . A similar result is obtained in the heat capacity measurements with a single sharp feature observed at 0 T which persists up to the highest measured field of 7 T with no observable change in temperature.

\section{\label{sec:con} Conclusions}

Our results show \TSOHO\ to be an interesting system to study the use of non-magnetic linker structural elements to form tunable magnetic sublattices. Studying the \textit{TM} = Mn, Ni and Co compounds, we first showed that all three crystallize in the \Pob\ space group with a complex network of highly distorted edge and corner sharing \textit{TM}O$_6$ octahedra. Simplistic quantifications of the structural triclinic \lq distortion\rq\ lead to the identification of the unit cell structure of \MSOHO\ being the least distorted while having the most distorted local environments for the \textit{TM} sites. This was in contrast to both \CSOHO\ and \NSOHO , which were relatively similar with larger triclinic distortions but more regular \textit{TM}O$_6$ octahedra. Magnetic susceptibility measurements showed all three compounds to have AFM like transitions at temperatures $<20$ K which Curie Weiss fitting revealed as moderately frustrated with frustration indexes around 3.5. Using neutron powder diffraction we confirmed these susceptibility features indicated the onset of long range AFM order with $k=(0,\frac{1}{2},\frac{1}{2})$ for \MSOHO\ and $k=(\frac{1}{2},\frac{1}{2},0)$ for both \CSOHO\ and \NSOHO .

Using Rietveld refinements and representational analysis, the low temperature magnetic structures were solved, showing an evolution from a relatively simple co-linear structure for \MSOHO\ to more complex canted structures for both \CSOHO\ and \NSOHO . Comparisons of the extracted magnetic moment to the effective moment of the Curie Weiss fits and calculated $TM^{2+}$ moment reveal disparities of $\sim 20$ and $\sim 15$ \%\ respectively, also potentially indicating frustrated physics. Field dependent heat capacity and magnetic susceptibility measurements, suggest possible metamagnetic transitions for both \MSOHO\ and \CSOHO . These latter observations suggest follow-up work using field dependent single crystal diffraction data to identify the metamagnetic states and inelastic neutron scattering to solve the spin-Hamiltonians. From there interesting physics may be realized through applying hydrostatic or uniaxial pressure to tune the structural and consequently magnetic properties. Additional work creating doping series may also be interesting as one tunes between the co-linear state of \MSOHO\ to the canted state of \CSOHO . Doing so may increase the frustration in these materials and lead to a quantum critical point as the material crosses from $k=(0,\frac{1}{2},\frac{1}{2})$ to $k=(\frac{1}{2},\frac{1}{2},0)$. Of further interest would be attempting to grow similar other structures using the SeO$_3$ non-magnetic linkers, or even larger such linkers, particularly if the dimensionality of the \textit{TM} sublattice could be decreased. Such work may lead to quantum effects as the magnetism is confined and the structural variability through \textit{TM} site allows for the different interaction pathways to be tuned.

\begin{acknowledgments}

We thank T. Ferreira for his assistance in Rietveld analysis of the neutron diffraction data. The part of the research conducted at ORNL’s Spallation Neutron Source was sponsored by the Scientific User Facilities Division, Office of Basic Energy Sciences (BES), U.S. Department of Energy (DOE). The research is supported by the U.S. DOE, BES, Materials Science and Engineering Division.

\end{acknowledgments}


\begin{thebibliography}{43}%
\makeatletter
\providecommand \@ifxundefined [1]{%
 \@ifx{#1\undefined}
}%
\providecommand \@ifnum [1]{%
 \ifnum #1\expandafter \@firstoftwo
 \else \expandafter \@secondoftwo
 \fi
}%
\providecommand \@ifx [1]{%
 \ifx #1\expandafter \@firstoftwo
 \else \expandafter \@secondoftwo
 \fi
}%
\providecommand \natexlab [1]{#1}%
\providecommand \enquote  [1]{``#1''}%
\providecommand \bibnamefont  [1]{#1}%
\providecommand \bibfnamefont [1]{#1}%
\providecommand \citenamefont [1]{#1}%
\providecommand \href@noop [0]{\@secondoftwo}%
\providecommand \href [0]{\begingroup \@sanitize@url \@href}%
\providecommand \@href[1]{\@@startlink{#1}\@@href}%
\providecommand \@@href[1]{\endgroup#1\@@endlink}%
\providecommand \@sanitize@url [0]{\catcode `\\12\catcode `\$12\catcode
  `\&12\catcode `\#12\catcode `\^12\catcode `\_12\catcode `\%12\relax}%
\providecommand \@@startlink[1]{}%
\providecommand \@@endlink[0]{}%
\providecommand \url  [0]{\begingroup\@sanitize@url \@url }%
\providecommand \@url [1]{\endgroup\@href {#1}{\urlprefix }}%
\providecommand \urlprefix  [0]{URL }%
\providecommand \Eprint [0]{\href }%
\providecommand \doibase [0]{http://dx.doi.org/}%
\providecommand \selectlanguage [0]{\@gobble}%
\providecommand \bibinfo  [0]{\@secondoftwo}%
\providecommand \bibfield  [0]{\@secondoftwo}%
\providecommand \translation [1]{[#1]}%
\providecommand \BibitemOpen [0]{}%
\providecommand \bibitemStop [0]{}%
\providecommand \bibitemNoStop [0]{.\EOS\space}%
\providecommand \EOS [0]{\spacefactor3000\relax}%
\providecommand \BibitemShut  [1]{\csname bibitem#1\endcsname}%
\let\auto@bib@innerbib\@empty
\bibitem [{\citenamefont {Chiu}\ \emph {et~al.}(2016)\citenamefont {Chiu},
  \citenamefont {Teo}, \citenamefont {Schnyder},\ and\ \citenamefont
  {Ryu}}]{Chiu2016}%
  \BibitemOpen
  \bibfield  {author} {\bibinfo {author} {\bibfnamefont {C.-K.}\ \bibnamefont
  {Chiu}}, \bibinfo {author} {\bibfnamefont {J.~C.~Y.}\ \bibnamefont {Teo}},
  \bibinfo {author} {\bibfnamefont {A.~P.}\ \bibnamefont {Schnyder}}, \ and\
  \bibinfo {author} {\bibfnamefont {S.}~\bibnamefont {Ryu}},\ }\href {\doibase
  10.1103/RevModPhys.88.035005} {\bibfield  {journal} {\bibinfo  {journal}
  {Rev. Mod. Phys.}\ }\textbf {\bibinfo {volume} {88}},\ \bibinfo {pages}
  {035005} (\bibinfo {year} {2016})}\BibitemShut {NoStop}%
\bibitem [{\citenamefont {Wen}(2017)}]{Wen2017}%
  \BibitemOpen
  \bibfield  {author} {\bibinfo {author} {\bibfnamefont {X.-G.}\ \bibnamefont
  {Wen}},\ }\href {\doibase 10.1103/RevModPhys.89.041004} {\bibfield  {journal}
  {\bibinfo  {journal} {Rev. Mod. Phys.}\ }\textbf {\bibinfo {volume} {89}},\
  \bibinfo {pages} {041004} (\bibinfo {year} {2017})}\BibitemShut {NoStop}%
\bibitem [{\citenamefont {Arovas}\ \emph {et~al.}(1984)\citenamefont {Arovas},
  \citenamefont {Schrieffer},\ and\ \citenamefont {Wilczek}}]{Arovas1984}%
  \BibitemOpen
  \bibfield  {author} {\bibinfo {author} {\bibfnamefont {D.}~\bibnamefont
  {Arovas}}, \bibinfo {author} {\bibfnamefont {J.}~\bibnamefont {Schrieffer}},
  \ and\ \bibinfo {author} {\bibfnamefont {F.}~\bibnamefont {Wilczek}},\ }\href
  {\doibase 10.1103/PhysRevLett.53.722} {\bibfield  {journal} {\bibinfo
  {journal} {Phys. Rev. Lett.}\ }\textbf {\bibinfo {volume} {53}},\ \bibinfo
  {pages} {722} (\bibinfo {year} {1984})}\BibitemShut {NoStop}%
\bibitem [{\citenamefont {Balents}(2010)}]{Balents2010}%
  \BibitemOpen
  \bibfield  {author} {\bibinfo {author} {\bibfnamefont {L.}~\bibnamefont
  {Balents}},\ }\href@noop {} {\bibfield  {journal} {\bibinfo  {journal}
  {Nature}\ }\textbf {\bibinfo {volume} {464}},\ \bibinfo {pages} {199}
  (\bibinfo {year} {2010})}\BibitemShut {NoStop}%
\bibitem [{\citenamefont {Castelnovo}\ \emph {et~al.}(2008)\citenamefont
  {Castelnovo}, \citenamefont {Moessner},\ and\ \citenamefont
  {Sondhi}}]{Castelnovo2008}%
  \BibitemOpen
  \bibfield  {author} {\bibinfo {author} {\bibfnamefont {C.}~\bibnamefont
  {Castelnovo}}, \bibinfo {author} {\bibfnamefont {R.}~\bibnamefont
  {Moessner}}, \ and\ \bibinfo {author} {\bibfnamefont {S.}~\bibnamefont
  {Sondhi}},\ }\href@noop {} {\bibfield  {journal} {\bibinfo  {journal}
  {Nature}\ }\textbf {\bibinfo {volume} {451}},\ \bibinfo {pages} {42}
  (\bibinfo {year} {2008})}\BibitemShut {NoStop}%
\bibitem [{\citenamefont {Gardner}\ \emph {et~al.}(2010)\citenamefont
  {Gardner}, \citenamefont {Gingras},\ and\ \citenamefont
  {Greedan}}]{Gardner2010}%
  \BibitemOpen
  \bibfield  {author} {\bibinfo {author} {\bibfnamefont {J.}~\bibnamefont
  {Gardner}}, \bibinfo {author} {\bibfnamefont {M.}~\bibnamefont {Gingras}}, \
  and\ \bibinfo {author} {\bibfnamefont {J.}~\bibnamefont {Greedan}},\
  }\href@noop {} {\bibfield  {journal} {\bibinfo  {journal} {Rev. Mod. Phys.}\
  }\textbf {\bibinfo {volume} {82}},\ \bibinfo {pages} {53} (\bibinfo {year}
  {2010})}\BibitemShut {NoStop}%
\bibitem [{\citenamefont {Ramirez}(1994)}]{Ramirez1994}%
  \BibitemOpen
  \bibfield  {author} {\bibinfo {author} {\bibfnamefont {A.}~\bibnamefont
  {Ramirez}},\ }\href@noop {} {\bibfield  {journal} {\bibinfo  {journal} {Annu.
  Rev. Mater. Sci.}\ }\textbf {\bibinfo {volume} {24}},\ \bibinfo {pages} {453}
  (\bibinfo {year} {1994})}\BibitemShut {NoStop}%
\bibitem [{\citenamefont {Derrida}\ \emph {et~al.}(1978)\citenamefont
  {Derrida}, \citenamefont {Vannimenus},\ and\ \citenamefont
  {Pomeau}}]{Derrida1978}%
  \BibitemOpen
  \bibfield  {author} {\bibinfo {author} {\bibfnamefont {B.}~\bibnamefont
  {Derrida}}, \bibinfo {author} {\bibfnamefont {J.}~\bibnamefont {Vannimenus}},
  \ and\ \bibinfo {author} {\bibfnamefont {Y.}~\bibnamefont {Pomeau}},\
  }\href@noop {} {\bibfield  {journal} {\bibinfo  {journal} {J. Phys. C: Solid
  State Phys.}\ }\textbf {\bibinfo {volume} {11}},\ \bibinfo {pages} {4749}
  (\bibinfo {year} {1978})}\BibitemShut {NoStop}%
\bibitem [{\citenamefont {Yan}\ \emph {et~al.}(2011)\citenamefont {Yan},
  \citenamefont {Huse},\ and\ \citenamefont {White}}]{Yan2011a}%
  \BibitemOpen
  \bibfield  {author} {\bibinfo {author} {\bibfnamefont {S.}~\bibnamefont
  {Yan}}, \bibinfo {author} {\bibfnamefont {D.}~\bibnamefont {Huse}}, \ and\
  \bibinfo {author} {\bibfnamefont {S.}~\bibnamefont {White}},\ }\href@noop {}
  {\bibfield  {journal} {\bibinfo  {journal} {Science}\ }\textbf {\bibinfo
  {volume} {332}},\ \bibinfo {pages} {1173} (\bibinfo {year}
  {2011})}\BibitemShut {NoStop}%
\bibitem [{\citenamefont {Mila}(2000)}]{Mila2000}%
  \BibitemOpen
  \bibfield  {author} {\bibinfo {author} {\bibfnamefont {F.}~\bibnamefont
  {Mila}},\ }\href@noop {} {\bibfield  {journal} {\bibinfo  {journal} {Eur. J.
  Phys.}\ }\textbf {\bibinfo {volume} {21}},\ \bibinfo {pages} {499} (\bibinfo
  {year} {2000})}\BibitemShut {NoStop}%
\bibitem [{\citenamefont {Han}\ \emph {et~al.}(2012)\citenamefont {Han},
  \citenamefont {Helton}, \citenamefont {Chu}, \citenamefont {Nocera},
  \citenamefont {Rodriguez-Rivera}, \citenamefont {Broholm},\ and\
  \citenamefont {Lee}}]{Han2012}%
  \BibitemOpen
  \bibfield  {author} {\bibinfo {author} {\bibfnamefont {T.-H.}\ \bibnamefont
  {Han}}, \bibinfo {author} {\bibfnamefont {J.}~\bibnamefont {Helton}},
  \bibinfo {author} {\bibfnamefont {S.}~\bibnamefont {Chu}}, \bibinfo {author}
  {\bibfnamefont {D.}~\bibnamefont {Nocera}}, \bibinfo {author} {\bibfnamefont
  {J.}~\bibnamefont {Rodriguez-Rivera}}, \bibinfo {author} {\bibfnamefont
  {C.}~\bibnamefont {Broholm}}, \ and\ \bibinfo {author} {\bibfnamefont
  {Y.}~\bibnamefont {Lee}},\ }\href@noop {} {\bibfield  {journal} {\bibinfo
  {journal} {Nature}\ }\textbf {\bibinfo {volume} {492}},\ \bibinfo {pages}
  {406} (\bibinfo {year} {2012})}\BibitemShut {NoStop}%
\bibitem [{\citenamefont {Starykh}(2015)}]{Starykh2015}%
  \BibitemOpen
  \bibfield  {author} {\bibinfo {author} {\bibfnamefont {O.}~\bibnamefont
  {Starykh}},\ }\href {\doibase 10.1088/0034-4885/78/5/052502} {\bibfield
  {journal} {\bibinfo  {journal} {Rep. Prog. Phys.}\ }\textbf {\bibinfo
  {volume} {78}},\ \bibinfo {pages} {052502} (\bibinfo {year}
  {2015})}\BibitemShut {NoStop}%
\bibitem [{\citenamefont {Willans}\ \emph {et~al.}(2010)\citenamefont
  {Willans}, \citenamefont {Chalker},\ and\ \citenamefont
  {Moessner}}]{Willans2010}%
  \BibitemOpen
  \bibfield  {author} {\bibinfo {author} {\bibfnamefont {A.}~\bibnamefont
  {Willans}}, \bibinfo {author} {\bibfnamefont {J.}~\bibnamefont {Chalker}}, \
  and\ \bibinfo {author} {\bibfnamefont {R.}~\bibnamefont {Moessner}},\
  }\href@noop {} {\bibfield  {journal} {\bibinfo  {journal} {Phys. Rev. Lett.}\
  }\textbf {\bibinfo {volume} {104}},\ \bibinfo {pages} {237203} (\bibinfo
  {year} {2010})}\BibitemShut {NoStop}%
\bibitem [{\citenamefont {Henley}(1989)}]{Henley1989}%
  \BibitemOpen
  \bibfield  {author} {\bibinfo {author} {\bibfnamefont {C.}~\bibnamefont
  {Henley}},\ }\href {\doibase 10.1103/PhysRevLett.62.2056} {\bibfield
  {journal} {\bibinfo  {journal} {Phys. Rev. Lett.}\ }\textbf {\bibinfo
  {volume} {62}},\ \bibinfo {pages} {2056} (\bibinfo {year}
  {1989})}\BibitemShut {NoStop}%
\bibitem [{\citenamefont {Sanjeewa}\ \emph {et~al.}(2016)\citenamefont
  {Sanjeewa}, \citenamefont {Garlea}, \citenamefont {McGuire}, \citenamefont
  {McMillen}, \citenamefont {Cao},\ and\ \citenamefont {Kolis}}]{Sanjeewa2016}%
  \BibitemOpen
  \bibfield  {author} {\bibinfo {author} {\bibfnamefont {L.~D.}\ \bibnamefont
  {Sanjeewa}}, \bibinfo {author} {\bibfnamefont {V.~O.}\ \bibnamefont
  {Garlea}}, \bibinfo {author} {\bibfnamefont {M.~A.}\ \bibnamefont {McGuire}},
  \bibinfo {author} {\bibfnamefont {C.~D.}\ \bibnamefont {McMillen}}, \bibinfo
  {author} {\bibfnamefont {H.}~\bibnamefont {Cao}}, \ and\ \bibinfo {author}
  {\bibfnamefont {J.~W.}\ \bibnamefont {Kolis}},\ }\href {\doibase
  10.1103/PhysRevB.93.224407} {\bibfield  {journal} {\bibinfo  {journal} {Phys.
  Rev. B}\ }\textbf {\bibinfo {volume} {93}},\ \bibinfo {pages} {224407}
  (\bibinfo {year} {2016})}\BibitemShut {NoStop}%
\bibitem [{\citenamefont {Garlea}\ \emph {et~al.}(2019)\citenamefont {Garlea},
  \citenamefont {Sanjeewa}, \citenamefont {McGuire}, \citenamefont {Batista},
  \citenamefont {Samarakoon}, \citenamefont {Graf}, \citenamefont {Winn},
  \citenamefont {Ye}, \citenamefont {Hoffmann},\ and\ \citenamefont
  {Kolis}}]{Garlea2019}%
  \BibitemOpen
  \bibfield  {author} {\bibinfo {author} {\bibfnamefont {V.}~\bibnamefont
  {Garlea}}, \bibinfo {author} {\bibfnamefont {L.}~\bibnamefont {Sanjeewa}},
  \bibinfo {author} {\bibfnamefont {M.}~\bibnamefont {McGuire}}, \bibinfo
  {author} {\bibfnamefont {C.}~\bibnamefont {Batista}}, \bibinfo {author}
  {\bibfnamefont {A.}~\bibnamefont {Samarakoon}}, \bibinfo {author}
  {\bibfnamefont {D.}~\bibnamefont {Graf}}, \bibinfo {author} {\bibfnamefont
  {B.}~\bibnamefont {Winn}}, \bibinfo {author} {\bibfnamefont {F.}~\bibnamefont
  {Ye}}, \bibinfo {author} {\bibfnamefont {C.}~\bibnamefont {Hoffmann}}, \ and\
  \bibinfo {author} {\bibfnamefont {J.}~\bibnamefont {Kolis}},\ }\href
  {\doibase 10.1103/PhysRevX.9.011038} {\bibfield  {journal} {\bibinfo
  {journal} {Phys. Rev. X}\ }\textbf {\bibinfo {volume} {9}},\ \bibinfo {pages}
  {011038} (\bibinfo {year} {2019})}\BibitemShut {NoStop}%
\bibitem [{\citenamefont {Zhang}\ \emph {et~al.}(2019)\citenamefont {Zhang},
  \citenamefont {Guo},\ and\ \citenamefont {He}}]{Zhang2019a}%
  \BibitemOpen
  \bibfield  {author} {\bibinfo {author} {\bibfnamefont {S.-Y.}\ \bibnamefont
  {Zhang}}, \bibinfo {author} {\bibfnamefont {W.-B.}\ \bibnamefont {Guo}}, \
  and\ \bibinfo {author} {\bibfnamefont {Z.-Z.}\ \bibnamefont {He}},\ }\href
  {\doibase 10.1039/C8DT03860K} {\bibfield  {journal} {\bibinfo  {journal}
  {Dalton Trans.}\ }\textbf {\bibinfo {volume} {48}},\ \bibinfo {pages} {65}
  (\bibinfo {year} {2019})}\BibitemShut {NoStop}%
\bibitem [{\citenamefont {Achary}\ \emph {et~al.}(2017)\citenamefont {Achary},
  \citenamefont {Bevara},\ and\ \citenamefont {Tyagi}}]{Achary2017}%
  \BibitemOpen
  \bibfield  {author} {\bibinfo {author} {\bibfnamefont {S.}~\bibnamefont
  {Achary}}, \bibinfo {author} {\bibfnamefont {S.}~\bibnamefont {Bevara}}, \
  and\ \bibinfo {author} {\bibfnamefont {A.}~\bibnamefont {Tyagi}},\
  }\href@noop {} {\bibfield  {journal} {\bibinfo  {journal} {Coord. Chem.
  Rev.}\ }\textbf {\bibinfo {volume} {340}},\ \bibinfo {pages} {266} (\bibinfo
  {year} {2017})}\BibitemShut {NoStop}%
\bibitem [{\citenamefont {Ren}\ \emph {et~al.}(2008)\citenamefont {Ren},
  \citenamefont {Liu}, \citenamefont {Cao}, \citenamefont {Zhao}, \citenamefont
  {Cao},\ and\ \citenamefont {Gao}}]{Ren2008a}%
  \BibitemOpen
  \bibfield  {author} {\bibinfo {author} {\bibfnamefont {Y.-H.}\ \bibnamefont
  {Ren}}, \bibinfo {author} {\bibfnamefont {S.-X.}\ \bibnamefont {Liu}},
  \bibinfo {author} {\bibfnamefont {R.-G.}\ \bibnamefont {Cao}}, \bibinfo
  {author} {\bibfnamefont {X.-Y.}\ \bibnamefont {Zhao}}, \bibinfo {author}
  {\bibfnamefont {J.-F.}\ \bibnamefont {Cao}}, \ and\ \bibinfo {author}
  {\bibfnamefont {C.-Y.}\ \bibnamefont {Gao}},\ }\href@noop {} {\bibfield
  {journal} {\bibinfo  {journal} {Inorg. Chem. Commun.}\ }\textbf {\bibinfo
  {volume} {11}},\ \bibinfo {pages} {1320} (\bibinfo {year}
  {2008})}\BibitemShut {NoStop}%
\bibitem [{\citenamefont {Harrison}(1999)}]{Harrison1999}%
  \BibitemOpen
  \bibfield  {author} {\bibinfo {author} {\bibfnamefont {W.}~\bibnamefont
  {Harrison}},\ }\href {\doibase 10.1107/S0108270199011099} {\bibfield
  {journal} {\bibinfo  {journal} {Acta Cryst. Sec. C}\ }\textbf {\bibinfo
  {volume} {55}},\ \bibinfo {pages} {1980} (\bibinfo {year}
  {1999})}\BibitemShut {NoStop}%
\bibitem [{\citenamefont {Mcmanus}\ \emph {et~al.}(1991)\citenamefont
  {Mcmanus}, \citenamefont {Harrison},\ and\ \citenamefont
  {Cheetham}}]{Mcmanus1991}%
  \BibitemOpen
  \bibfield  {author} {\bibinfo {author} {\bibfnamefont {A.}~\bibnamefont
  {Mcmanus}}, \bibinfo {author} {\bibfnamefont {W.}~\bibnamefont {Harrison}}, \
  and\ \bibinfo {author} {\bibfnamefont {A.}~\bibnamefont {Cheetham}},\ }\href
  {\doibase https://doi.org/10.1016/0022-4596(91)90333-D} {\bibfield  {journal}
  {\bibinfo  {journal} {J. Solid State Chem.}\ }\textbf {\bibinfo {volume}
  {92}},\ \bibinfo {pages} {253 } (\bibinfo {year} {1991})}\BibitemShut
  {NoStop}%
\bibitem [{\citenamefont {Wildner}(1991)}]{Wildner1991}%
  \BibitemOpen
  \bibfield  {author} {\bibinfo {author} {\bibfnamefont {M.}~\bibnamefont
  {Wildner}},\ }\href {\doibase 10.1007/BF00811457} {\bibfield  {journal}
  {\bibinfo  {journal} {Monatshefte f{\"u}r Chemie / Chemical Monthly}\
  }\textbf {\bibinfo {volume} {122}},\ \bibinfo {pages} {585} (\bibinfo {year}
  {1991})}\BibitemShut {NoStop}%
\bibitem [{\citenamefont {Larranaga}\ \emph {et~al.}(2002)\citenamefont
  {Larranaga}, \citenamefont {Mesa}, \citenamefont {Pizarro}, \citenamefont
  {Olazcuaga}, \citenamefont {Arriortua},\ and\ \citenamefont
  {Rojo}}]{Larranaga2002}%
  \BibitemOpen
  \bibfield  {author} {\bibinfo {author} {\bibfnamefont {A.}~\bibnamefont
  {Larranaga}}, \bibinfo {author} {\bibfnamefont {J.}~\bibnamefont {Mesa}},
  \bibinfo {author} {\bibfnamefont {J.}~\bibnamefont {Pizarro}}, \bibinfo
  {author} {\bibfnamefont {R.}~\bibnamefont {Olazcuaga}}, \bibinfo {author}
  {\bibfnamefont {M.}~\bibnamefont {Arriortua}}, \ and\ \bibinfo {author}
  {\bibfnamefont {T.}~\bibnamefont {Rojo}},\ }\href {\doibase 10.1039/B206515K}
  {\bibfield  {journal} {\bibinfo  {journal} {J. Chem. Soc.{,} Dalton Trans.}\
  ,\ \bibinfo {pages} {3447}} (\bibinfo {year} {2002})}\BibitemShut {NoStop}%
\bibitem [{\citenamefont {Calder}\ \emph {et~al.}(2018)\citenamefont {Calder},
  \citenamefont {An}, \citenamefont {Boehler}, \citenamefont {Dela~Cruz},
  \citenamefont {Frontzek}, \citenamefont {Guthrie}, \citenamefont {Haberl},
  \citenamefont {Huq}, \citenamefont {Kimber}, \citenamefont {Liu} \emph
  {et~al.}}]{Calder2018}%
  \BibitemOpen
  \bibfield  {author} {\bibinfo {author} {\bibfnamefont {S.}~\bibnamefont
  {Calder}}, \bibinfo {author} {\bibfnamefont {K.}~\bibnamefont {An}}, \bibinfo
  {author} {\bibfnamefont {R.}~\bibnamefont {Boehler}}, \bibinfo {author}
  {\bibfnamefont {C.}~\bibnamefont {Dela~Cruz}}, \bibinfo {author}
  {\bibfnamefont {M.}~\bibnamefont {Frontzek}}, \bibinfo {author}
  {\bibfnamefont {M.}~\bibnamefont {Guthrie}}, \bibinfo {author} {\bibfnamefont
  {B.}~\bibnamefont {Haberl}}, \bibinfo {author} {\bibfnamefont
  {A.}~\bibnamefont {Huq}}, \bibinfo {author} {\bibfnamefont {S.~A.}\
  \bibnamefont {Kimber}}, \bibinfo {author} {\bibfnamefont {J.}~\bibnamefont
  {Liu}},  \emph {et~al.},\ }\href@noop {} {\bibfield  {journal} {\bibinfo
  {journal} {Rev. Sci. Inst.}\ }\textbf {\bibinfo {volume} {89}},\ \bibinfo
  {pages} {092701} (\bibinfo {year} {2018})}\BibitemShut {NoStop}%
\bibitem [{\citenamefont
  {Rodr{\'{\i}}guez-Carvajal}(1993)}]{Rodriguez-Carvajal1993}%
  \BibitemOpen
  \bibfield  {author} {\bibinfo {author} {\bibfnamefont {J.}~\bibnamefont
  {Rodr{\'{\i}}guez-Carvajal}},\ }\href {\doibase 10.1016/0921-4526(93)90108-I}
  {\bibfield  {journal} {\bibinfo  {journal} {Phys. B}\ }\textbf {\bibinfo
  {volume} {192}},\ \bibinfo {pages} {55} (\bibinfo {year} {1993})}\BibitemShut
  {NoStop}%
\bibitem [{\citenamefont {Toby}\ and\ \citenamefont
  {Von~Dreele}(2013)}]{Toby2013}%
  \BibitemOpen
  \bibfield  {author} {\bibinfo {author} {\bibfnamefont {B.}~\bibnamefont
  {Toby}}\ and\ \bibinfo {author} {\bibfnamefont {R.}~\bibnamefont
  {Von~Dreele}},\ }\href {\doibase 10.1107/S0021889813003531} {\bibfield
  {journal} {\bibinfo  {journal} {J. Appl. Crystallogr.}\ }\textbf {\bibinfo
  {volume} {46}},\ \bibinfo {pages} {544} (\bibinfo {year} {2013})}\BibitemShut
  {NoStop}%
\bibitem [{\citenamefont {Wills}(2000)}]{Wills2000}%
  \BibitemOpen
  \bibfield  {author} {\bibinfo {author} {\bibfnamefont {A.}~\bibnamefont
  {Wills}},\ }\href {\doibase https://doi.org/10.1016/S0921-4526(99)01722-6}
  {\bibfield  {journal} {\bibinfo  {journal} {Phys. B}\ }\textbf {\bibinfo
  {volume} {276-278}},\ \bibinfo {pages} {680 } (\bibinfo {year}
  {2000})}\BibitemShut {NoStop}%
\bibitem [{\citenamefont {Aroyo}\ \emph
  {et~al.}(2006{\natexlab{a}})\citenamefont {Aroyo}, \citenamefont
  {Perez-Mato}, \citenamefont {Capillas}, \citenamefont {Kroumova},
  \citenamefont {Ivantchev}, \citenamefont {Madariaga}, \citenamefont {Kirov},\
  and\ \citenamefont {Wondratschek}}]{Aroyo2006a}%
  \BibitemOpen
  \bibfield  {author} {\bibinfo {author} {\bibfnamefont {M.}~\bibnamefont
  {Aroyo}}, \bibinfo {author} {\bibfnamefont {J.}~\bibnamefont {Perez-Mato}},
  \bibinfo {author} {\bibfnamefont {C.}~\bibnamefont {Capillas}}, \bibinfo
  {author} {\bibfnamefont {E.}~\bibnamefont {Kroumova}}, \bibinfo {author}
  {\bibfnamefont {S.}~\bibnamefont {Ivantchev}}, \bibinfo {author}
  {\bibfnamefont {G.}~\bibnamefont {Madariaga}}, \bibinfo {author}
  {\bibfnamefont {A.}~\bibnamefont {Kirov}}, \ and\ \bibinfo {author}
  {\bibfnamefont {H.}~\bibnamefont {Wondratschek}},\ }\href {\doibase
  10.1524/zkri.2006.221.1.15} {\bibfield  {journal} {\bibinfo  {journal} {Z.
  KRIST.}\ }\textbf {\bibinfo {volume} {221}},\ \bibinfo {pages} {15} (\bibinfo
  {year} {2006}{\natexlab{a}})},\ \bibinfo {note} {workshop on Crystallography
  at the Start of the 21st Century - Mathematical and Symmetry Aspects,
  Budapest, HUNGARY, AUG 24-26, 2004-2005}\BibitemShut {NoStop}%
\bibitem [{\citenamefont {Aroyo}\ \emph
  {et~al.}(2006{\natexlab{b}})\citenamefont {Aroyo}, \citenamefont {Kirov},
  \citenamefont {Capillas}, \citenamefont {Perez-Mato},\ and\ \citenamefont
  {Wondratschek}}]{Aroyo2006b}%
  \BibitemOpen
  \bibfield  {author} {\bibinfo {author} {\bibfnamefont {M.}~\bibnamefont
  {Aroyo}}, \bibinfo {author} {\bibfnamefont {A.}~\bibnamefont {Kirov}},
  \bibinfo {author} {\bibfnamefont {C.}~\bibnamefont {Capillas}}, \bibinfo
  {author} {\bibfnamefont {J.}~\bibnamefont {Perez-Mato}}, \ and\ \bibinfo
  {author} {\bibfnamefont {H.}~\bibnamefont {Wondratschek}},\ }\href {\doibase
  10.1107/S0108767305040286} {\bibfield  {journal} {\bibinfo  {journal} {Acta
  Crystallogr. Sec. A}\ }\textbf {\bibinfo {volume} {62}},\ \bibinfo {pages}
  {115} (\bibinfo {year} {2006}{\natexlab{b}})}\BibitemShut {NoStop}%
\bibitem [{\citenamefont {Aroyo}\ \emph {et~al.}(2011)\citenamefont {Aroyo},
  \citenamefont {Perez-Mato}, \citenamefont {Orobengoa}, \citenamefont {Tasci},
  \citenamefont {De~La~Flor},\ and\ \citenamefont {Kirov}}]{Aroyo2011}%
  \BibitemOpen
  \bibfield  {author} {\bibinfo {author} {\bibfnamefont {M.}~\bibnamefont
  {Aroyo}}, \bibinfo {author} {\bibfnamefont {J.}~\bibnamefont {Perez-Mato}},
  \bibinfo {author} {\bibfnamefont {D.}~\bibnamefont {Orobengoa}}, \bibinfo
  {author} {\bibfnamefont {E.}~\bibnamefont {Tasci}}, \bibinfo {author}
  {\bibfnamefont {G.}~\bibnamefont {De~La~Flor}}, \ and\ \bibinfo {author}
  {\bibfnamefont {A.}~\bibnamefont {Kirov}},\ }\href
  {https://www2.scopus.com/inward/record.uri?eid=2-s2.0-80955140447&partnerID=40&md5=488772b9e21d2636a3952f66ae80ae84}
  {\bibfield  {journal} {\bibinfo  {journal} {Bulg. Chem. Commun.}\ }\textbf
  {\bibinfo {volume} {43}},\ \bibinfo {pages} {183} (\bibinfo {year}
  {2011})}\BibitemShut {NoStop}%
\bibitem [{\citenamefont {Momma}\ and\ \citenamefont
  {Izumi}(2011)}]{Momma2011}%
  \BibitemOpen
  \bibfield  {author} {\bibinfo {author} {\bibfnamefont {K.}~\bibnamefont
  {Momma}}\ and\ \bibinfo {author} {\bibfnamefont {F.}~\bibnamefont {Izumi}},\
  }\href@noop {} {\bibfield  {journal} {\bibinfo  {journal} {J. Appl.
  Crysatllogr.}\ }\textbf {\bibinfo {volume} {44}},\ \bibinfo {pages} {1272}
  (\bibinfo {year} {2011})}\BibitemShut {NoStop}%
\bibitem [{\citenamefont {Blaha}\ \emph {et~al.}(2018)\citenamefont {Blaha},
  \citenamefont {Schwarz}, \citenamefont {Madsen}, \citenamefont {Kvasnicka},
  \citenamefont {Luitz}, \citenamefont {Laskowski}, \citenamefont {Tran},\ and\
  \citenamefont {Marks}}]{Blaha2018}%
  \BibitemOpen
  \bibfield  {author} {\bibinfo {author} {\bibfnamefont {P.}~\bibnamefont
  {Blaha}}, \bibinfo {author} {\bibfnamefont {K.}~\bibnamefont {Schwarz}},
  \bibinfo {author} {\bibfnamefont {G.~K.~H.}\ \bibnamefont {Madsen}}, \bibinfo
  {author} {\bibfnamefont {D.}~\bibnamefont {Kvasnicka}}, \bibinfo {author}
  {\bibfnamefont {J.}~\bibnamefont {Luitz}}, \bibinfo {author} {\bibfnamefont
  {R.}~\bibnamefont {Laskowski}}, \bibinfo {author} {\bibfnamefont
  {F.}~\bibnamefont {Tran}}, \ and\ \bibinfo {author} {\bibfnamefont {L.~D.}\
  \bibnamefont {Marks}},\ }\href@noop {} {\emph {\bibinfo {title} {WIEN2k, An
  Augmented Plane Wave + Local Orbitals Program for Calculating Crystal
  Properties}}}\ (\bibinfo  {publisher} {Karlheinz Schwarz},\ \bibinfo
  {address} {Techn. Universität Wien, Austria},\ \bibinfo {year}
  {2018})\BibitemShut {NoStop}%
\bibitem [{\citenamefont {Perdew}\ \emph {et~al.}(1996)\citenamefont {Perdew},
  \citenamefont {Burke},\ and\ \citenamefont {Ernzerhof}}]{Perdew1996}%
  \BibitemOpen
  \bibfield  {author} {\bibinfo {author} {\bibfnamefont {J.}~\bibnamefont
  {Perdew}}, \bibinfo {author} {\bibfnamefont {K.}~\bibnamefont {Burke}}, \
  and\ \bibinfo {author} {\bibfnamefont {M.}~\bibnamefont {Ernzerhof}},\
  }\href@noop {} {\bibfield  {journal} {\bibinfo  {journal} {Phys. Rev. Lett.}\
  }\textbf {\bibinfo {volume} {77}},\ \bibinfo {pages} {3865} (\bibinfo {year}
  {1996})}\BibitemShut {NoStop}%
\bibitem [{\citenamefont {Shannon}(1976)}]{Shannon1976}%
  \BibitemOpen
  \bibfield  {author} {\bibinfo {author} {\bibfnamefont {R.}~\bibnamefont
  {Shannon}},\ }\href {\doibase 10.1107/S0567739476001551} {\bibfield
  {journal} {\bibinfo  {journal} {Acta Crystallogr.}\ }\textbf {\bibinfo
  {volume} {32}},\ \bibinfo {pages} {751} (\bibinfo {year} {1976})}\BibitemShut
  {NoStop}%
\bibitem [{\citenamefont {Foadi}\ and\ \citenamefont
  {Evans}(2011)}]{Foadi2011}%
  \BibitemOpen
  \bibfield  {author} {\bibinfo {author} {\bibfnamefont {J.}~\bibnamefont
  {Foadi}}\ and\ \bibinfo {author} {\bibfnamefont {G.}~\bibnamefont {Evans}},\
  }\href {\doibase 10.1107/S0108767310044296} {\bibfield  {journal} {\bibinfo
  {journal} {Acta Crystallogr.}\ }\textbf {\bibinfo {volume} {67}},\ \bibinfo
  {pages} {93} (\bibinfo {year} {2011})}\BibitemShut {NoStop}%
\bibitem [{\citenamefont {Geertsma}(1990)}]{Geertsma1990}%
  \BibitemOpen
  \bibfield  {author} {\bibinfo {author} {\bibfnamefont {W.}~\bibnamefont
  {Geertsma}},\ }\href {\doibase https://doi.org/10.1016/0921-4526(90)90812-9}
  {\bibfield  {journal} {\bibinfo  {journal} {Phys. B}\ }\textbf {\bibinfo
  {volume} {164}},\ \bibinfo {pages} {241 } (\bibinfo {year}
  {1990})}\BibitemShut {NoStop}%
\bibitem [{\citenamefont {Choudhury}\ \emph {et~al.}(2002)\citenamefont
  {Choudhury}, \citenamefont {Kumar~D},\ and\ \citenamefont
  {Rao}}]{Choudhury2002}%
  \BibitemOpen
  \bibfield  {author} {\bibinfo {author} {\bibfnamefont {A.}~\bibnamefont
  {Choudhury}}, \bibinfo {author} {\bibfnamefont {U.}~\bibnamefont {Kumar~D}},
  \ and\ \bibinfo {author} {\bibfnamefont {C.}~\bibnamefont {Rao}},\ }\href
  {\doibase 10.1002/1521-3773(20020104)41:1<158::AID-ANIE158>3.0.CO;2-\#}
  {\bibfield  {journal} {\bibinfo  {journal} {Angew. Chem., Int. Ed.}\ }\textbf
  {\bibinfo {volume} {41}},\ \bibinfo {pages} {158} (\bibinfo {year}
  {2002})}\BibitemShut {NoStop}%
\bibitem [{\citenamefont {Kovrugin}\ \emph {et~al.}(2018)\citenamefont
  {Kovrugin}, \citenamefont {Colmont}, \citenamefont {Siidra}, \citenamefont
  {Charkin}, \citenamefont {Aliev}, \citenamefont {Krivovichev},\ and\
  \citenamefont {Mentre}}]{Kovrugin2018}%
  \BibitemOpen
  \bibfield  {author} {\bibinfo {author} {\bibfnamefont {V.}~\bibnamefont
  {Kovrugin}}, \bibinfo {author} {\bibfnamefont {M.}~\bibnamefont {Colmont}},
  \bibinfo {author} {\bibfnamefont {O.}~\bibnamefont {Siidra}}, \bibinfo
  {author} {\bibfnamefont {D.}~\bibnamefont {Charkin}}, \bibinfo {author}
  {\bibfnamefont {A.}~\bibnamefont {Aliev}}, \bibinfo {author} {\bibfnamefont
  {S.}~\bibnamefont {Krivovichev}}, \ and\ \bibinfo {author} {\bibfnamefont
  {O.}~\bibnamefont {Mentre}},\ }\href {\doibase 10.1515/zkri-2018-2088}
  {\bibfield  {journal} {\bibinfo  {journal} {Z. Kristallogr. - Cryst. Mater.}\
  }\textbf {\bibinfo {volume} {234}} (\bibinfo {year} {2018}),\
  10.1515/zkri-2018-2088}\BibitemShut {NoStop}%
\bibitem [{\citenamefont {Dun}\ \emph {et~al.}(2015)\citenamefont {Dun},
  \citenamefont {Li}, \citenamefont {Freitas}, \citenamefont {Arrighi},
  \citenamefont {Dela~Cruz}, \citenamefont {Lee}, \citenamefont {Choi},
  \citenamefont {Cao}, \citenamefont {Silverstein}, \citenamefont {Wiebe},
  \citenamefont {Cheng},\ and\ \citenamefont {Zhou}}]{Dun2015}%
  \BibitemOpen
  \bibfield  {author} {\bibinfo {author} {\bibfnamefont {Z.~L.}\ \bibnamefont
  {Dun}}, \bibinfo {author} {\bibfnamefont {X.}~\bibnamefont {Li}}, \bibinfo
  {author} {\bibfnamefont {R.~S.}\ \bibnamefont {Freitas}}, \bibinfo {author}
  {\bibfnamefont {E.}~\bibnamefont {Arrighi}}, \bibinfo {author} {\bibfnamefont
  {C.~R.}\ \bibnamefont {Dela~Cruz}}, \bibinfo {author} {\bibfnamefont
  {M.}~\bibnamefont {Lee}}, \bibinfo {author} {\bibfnamefont {E.~S.}\
  \bibnamefont {Choi}}, \bibinfo {author} {\bibfnamefont {H.~B.}\ \bibnamefont
  {Cao}}, \bibinfo {author} {\bibfnamefont {H.~J.}\ \bibnamefont
  {Silverstein}}, \bibinfo {author} {\bibfnamefont {C.~R.}\ \bibnamefont
  {Wiebe}}, \bibinfo {author} {\bibfnamefont {J.~G.}\ \bibnamefont {Cheng}}, \
  and\ \bibinfo {author} {\bibfnamefont {H.~D.}\ \bibnamefont {Zhou}},\ }\href
  {\doibase 10.1103/PhysRevB.92.140407} {\bibfield  {journal} {\bibinfo
  {journal} {Phys. Rev. B}\ }\textbf {\bibinfo {volume} {92}},\ \bibinfo
  {pages} {140407} (\bibinfo {year} {2015})}\BibitemShut {NoStop}%
\bibitem [{\citenamefont {Li}\ \emph {et~al.}(2014)\citenamefont {Li},
  \citenamefont {Li}, \citenamefont {Matsubayashi}, \citenamefont {Sato},
  \citenamefont {Jin}, \citenamefont {Uwatoko}, \citenamefont {Kawae},
  \citenamefont {Hallas}, \citenamefont {Wiebe}, \citenamefont {Arevalo-Lopez},
  \citenamefont {Attfield}, \citenamefont {Gardner}, \citenamefont {Freitas},
  \citenamefont {Zhou},\ and\ \citenamefont {Cheng}}]{Li2014}%
  \BibitemOpen
  \bibfield  {author} {\bibinfo {author} {\bibfnamefont {X.}~\bibnamefont
  {Li}}, \bibinfo {author} {\bibfnamefont {W.~M.}\ \bibnamefont {Li}}, \bibinfo
  {author} {\bibfnamefont {K.}~\bibnamefont {Matsubayashi}}, \bibinfo {author}
  {\bibfnamefont {Y.}~\bibnamefont {Sato}}, \bibinfo {author} {\bibfnamefont
  {C.~Q.}\ \bibnamefont {Jin}}, \bibinfo {author} {\bibfnamefont
  {Y.}~\bibnamefont {Uwatoko}}, \bibinfo {author} {\bibfnamefont
  {T.}~\bibnamefont {Kawae}}, \bibinfo {author} {\bibfnamefont {A.~M.}\
  \bibnamefont {Hallas}}, \bibinfo {author} {\bibfnamefont {C.~R.}\
  \bibnamefont {Wiebe}}, \bibinfo {author} {\bibfnamefont {A.~M.}\ \bibnamefont
  {Arevalo-Lopez}}, \bibinfo {author} {\bibfnamefont {J.~P.}\ \bibnamefont
  {Attfield}}, \bibinfo {author} {\bibfnamefont {J.~S.}\ \bibnamefont
  {Gardner}}, \bibinfo {author} {\bibfnamefont {R.~S.}\ \bibnamefont
  {Freitas}}, \bibinfo {author} {\bibfnamefont {H.~D.}\ \bibnamefont {Zhou}}, \
  and\ \bibinfo {author} {\bibfnamefont {J.-G.}\ \bibnamefont {Cheng}},\ }\href
  {\doibase 10.1103/PhysRevB.89.064409} {\bibfield  {journal} {\bibinfo
  {journal} {Phys. Rev. B}\ }\textbf {\bibinfo {volume} {89}},\ \bibinfo
  {pages} {064409} (\bibinfo {year} {2014})}\BibitemShut {NoStop}%
\bibitem [{\citenamefont {Chen}\ \emph {et~al.}(2019)\citenamefont {Chen},
  \citenamefont {Wang}, \citenamefont {Yan}, \citenamefont {Parker},
  \citenamefont {Zhou}, \citenamefont {Uwatoko},\ and\ \citenamefont
  {Cheng}}]{Chen2019}%
  \BibitemOpen
  \bibfield  {author} {\bibinfo {author} {\bibfnamefont {K.}~\bibnamefont
  {Chen}}, \bibinfo {author} {\bibfnamefont {B.}~\bibnamefont {Wang}}, \bibinfo
  {author} {\bibfnamefont {J.-Q.}\ \bibnamefont {Yan}}, \bibinfo {author}
  {\bibfnamefont {D.}~\bibnamefont {Parker}}, \bibinfo {author} {\bibfnamefont
  {J.-S.}\ \bibnamefont {Zhou}}, \bibinfo {author} {\bibfnamefont
  {Y.}~\bibnamefont {Uwatoko}}, \ and\ \bibinfo {author} {\bibfnamefont
  {J.-G.}\ \bibnamefont {Cheng}},\ }\href@noop {} {\bibfield  {journal}
  {\bibinfo  {journal} {Phys. Rev. Mat.}\ }\textbf {\bibinfo {volume} {3}},\
  \bibinfo {pages} {094201} (\bibinfo {year} {2019})}\BibitemShut {NoStop}%
\bibitem [{\citenamefont {Pokharel}\ \emph {et~al.}(2018)\citenamefont
  {Pokharel}, \citenamefont {May}, \citenamefont {Parker}, \citenamefont
  {Calder}, \citenamefont {Ehlers}, \citenamefont {Huq}, \citenamefont
  {Kimber}, \citenamefont {Arachchige}, \citenamefont {Poudel}, \citenamefont
  {McGuire} \emph {et~al.}}]{Pokharel2018}%
  \BibitemOpen
  \bibfield  {author} {\bibinfo {author} {\bibfnamefont {G.}~\bibnamefont
  {Pokharel}}, \bibinfo {author} {\bibfnamefont {A.}~\bibnamefont {May}},
  \bibinfo {author} {\bibfnamefont {D.}~\bibnamefont {Parker}}, \bibinfo
  {author} {\bibfnamefont {S.}~\bibnamefont {Calder}}, \bibinfo {author}
  {\bibfnamefont {G.}~\bibnamefont {Ehlers}}, \bibinfo {author} {\bibfnamefont
  {A.}~\bibnamefont {Huq}}, \bibinfo {author} {\bibfnamefont {S.}~\bibnamefont
  {Kimber}}, \bibinfo {author} {\bibfnamefont {H.}~\bibnamefont {Arachchige}},
  \bibinfo {author} {\bibfnamefont {L.}~\bibnamefont {Poudel}}, \bibinfo
  {author} {\bibfnamefont {M.}~\bibnamefont {McGuire}},  \emph {et~al.},\
  }\href@noop {} {\bibfield  {journal} {\bibinfo  {journal} {Phys. Rev. B}\
  }\textbf {\bibinfo {volume} {97}},\ \bibinfo {pages} {134117} (\bibinfo
  {year} {2018})}\BibitemShut {NoStop}%
\bibitem [{\citenamefont {Taddei}\ \emph {et~al.}(2019)\citenamefont {Taddei},
  \citenamefont {Sanjeewa}, \citenamefont {Kolis}, \citenamefont {Sefat},
  \citenamefont {de~la Cruz},\ and\ \citenamefont {Pajerowski}}]{Taddei2019}%
  \BibitemOpen
  \bibfield  {author} {\bibinfo {author} {\bibfnamefont {K.}~\bibnamefont
  {Taddei}}, \bibinfo {author} {\bibfnamefont {L.}~\bibnamefont {Sanjeewa}},
  \bibinfo {author} {\bibfnamefont {J.}~\bibnamefont {Kolis}}, \bibinfo
  {author} {\bibfnamefont {A.}~\bibnamefont {Sefat}}, \bibinfo {author}
  {\bibfnamefont {C.}~\bibnamefont {de~la Cruz}}, \ and\ \bibinfo {author}
  {\bibfnamefont {D.}~\bibnamefont {Pajerowski}},\ }\href {\doibase
  10.1103/PhysRevMaterials.3.014405} {\bibfield  {journal} {\bibinfo  {journal}
  {Phys. Rev. Materials}\ }\textbf {\bibinfo {volume} {3}},\ \bibinfo {pages}
  {014405} (\bibinfo {year} {2019})}\BibitemShut {NoStop}%
\end{thebibliography}

%

\end{document}